\documentclass[aps,amsmath,amssymb,superscriptaddress,prb,reprint,twocolumn,longbibliography]{revtex4-1}
\usepackage{graphicx}
\usepackage{dcolumn}
\usepackage{braket,bm}

\usepackage{mathptmx}

\begin{document}

\preprint{APS/123-QED}

\title{Nonlinear optical Hall effect of few-layered NbSe$_2$}
\author{Ren Habara}
\affiliation{Department of Nanotechnology for Sustainable Energy, School
of Science and Technology, Kwansei Gakuin University, Gakuen 2-1, Sanda 669-1337, Japan}
\author{Katsunori Wakabayashi}
\affiliation{Department of Nanotechnology for Sustainable Energy, School
of Science and Technology, Kwansei Gakuin University, Gakuen 2-1, Sanda 669-1337, Japan}
\affiliation{National Institute for Materials Science (NIMS), Namiki 1-1, Tsukuba 305-0044, Japan}
\affiliation{Center for Spintronics Research Network (CSRN), Osaka
University, Toyonaka 560-8531, Japan}

\date{\today}

\begin{abstract}
NbSe$_2$ is one of metallic two-dimensional (2D) transition-metal
dichalcogenide (TMDC) materials. Because of broken crystal inversion symmetry,
 large spin splitting is induced by Ising-type spin-orbit 
coupling in odd-number-layered NbSe$_2$, but absent for
even-number-layered NbSe$_2$ with the inversion symmetry.
In this paper, we numerically calculate nonlinear optical charge and
 spin Hall conductivities of 
few-layered NbSe$_2$ based on an effective tight-binding model which
includes $d_{z^2}$, $d_{x^2-y^2}$ and $d_{xy}$ orbitals of Nb atom.
We show that the nonlinear optical Hall
conductivity for second harmonic generation (SHG) process has nonvanishing
value in odd-number-layered NbSe$_2$. Also, we provide 
nonlinear optical selection rule in
 few-layered NbSe$_2$ and their polarization dependences. 
In further, for even-number-layered case, the nonlinear optical Hall
 currents can be generated by applying electric fields which breaks
 inversion symmetry.  
We also discuss that the nonlinear optical Hall effect is expected to
 occur in TMDC materials in general.
Thus, our results will serve to design potential
 opt-spintronics devices based on 2D materials to generate the spin Hall
 current by SHG.
\end{abstract}

\maketitle

\section{Introduction}\label{sec:level1}
Transition metal dichalcogenide (TMDC) with a chemical formula MX$_2$ (M=Mo, W, Nb, Ta; X=S, Se) is a new class of
two-dimensional (2D) electronic systems, which
provides the platform to design functional opt-electronic
devices.~\cite{Mak2010, Sple2010, Ton2012, Gut2013, Zhao2013,
  Ciarrocchi2018, Samadi2018, Thakar2020, Wu2021}
Owing to weak van der Waals forces between layers, bulk TMDC can be
easily exfoliated into monolayer.~\cite{Novoselov2005, Desai2016, Lin2016,
  Yu2018, Wang2021}
In monolayer MoS$_2$ and WSe$_2$,
the valley-dependent optical excitation~\cite{Zeng, Mak, Cao, Yu} and
intrinsic spin Hall effect (SHE)~\cite{Sinova, Hai, Kato, Wun} 
have been reported.
In further, nonlinear optical effect such as 
second-harmonic generation (SHG),~\cite{Wen2019, Lucas2021,
  Taghizadeh2019, Trolle2014, Kumar2013, Moss1987, Ghahramani1991,
  Malard2013, Seyler2015, Rashkeev1998, Rashkeev2001, Sharma2004,
  Leitsmann2005} sum-frequency generation (SFG),~\cite{Shen1989,
  Shen2020, Shultz2000, Richmond2002, Shen2006, Gopalakrishnan2006,
  Ishiyama2014} third-harmonic generation (THG),~\cite{Wen2019,
  Tsang1995, Jeff1998, Stock2020} high-harmonic generation
(HHG)~\cite{Wen2019, Dromey2006, Winterfeldt2008, Vampa2014} and
two-photon absorption~\cite{Wen2019, Pawlicki2009, Rumi2010, Wang2020} has been extensively studied. 
In general, the nonlinear optical effect is sensitive to crystal
symmetries and phase-matching conditions between incident light and
light-induced electric polarization wave. 
However,
the phase-matching conditions are not necessary for
nonlinear optical effect in atomically-thin 2D materials,
because their thickness is much smaller than the light
wavelength.~\cite{Wen2019, Fryett2017, He2021} 
Thus, in atomically-thin 2D materials such as TMDCs,
the nonlinear optical effect strongly depends on the crystal symmetry.
In addition, 
it has been recently discussed that second-order anomalous transport
phenomena in the absence of magnetic field can be induced by uniaxial
strain in metallic TMDC, i.e., nonlinear Hall effect, which has a
relation with Berry curvature dipole.~\cite{Xiao2020, Zhou2020, Ma2019,
Sodemann2015}

NbSe$_2$ is metallic TMDC which shows superconducting phase transition 
at low temperatures.~\cite{Wilson2001, Kim, He, Xi, Sohn, Anikin2020, Lian2017}
In this material, AB-stacking structure is most stable in nature
and has different crystal symmetries for even and odd number of layers.
In even-number-layered NbSe$_2$, the crystal structure has a space group
D$_{3d}$, which respects to inversion and out-of-plane mirror
symmetries.
On the other hand, odd-number-layered NbSe$_2$ has a space group
D$_{3h}$, which possesses out-of-plane mirror symmetry, but no spacial
inversion symmetry.
Because of the broken inversion symmetry and a strong
 spin-orbit
coupling (SOC) field of Nb atoms
 in odd-number-layered
NbSe$_2$, it possesses Ising-type SOC,~\cite{He, Xi, Sohn, Lu, Saito,
  Zhou, Baw} i.e., an effective Zeeman field that locks electron spins
to out-of-plane directions by in-plane momentum and 
causes larger spin splitting in the energy band structures leading to 
unconventional topological spin properties.
In actual, we have analyzed the linear optical properties of monolayer
NbSe$_2$ using Kubo formula based on an effective tight-binding model
(TBM) and shown that the spin Hall current can be induced by
irradiating visible light owing to its finite spin Berry curvature.~\cite{Ren2021}

In this paper, we extend our theoretical analysis to nonlinear optical spin and
charge Hall conductivities of SHG process for
few-layered NbSe$_2$.
Here, we have employed the effective TBM to describe
the electronic structures of few-layered NbSe$_2$, 
where the electron hoppings among 
$d_{z^2}$, $d_{x^2-y^2}$ and $d_{xy}$ orbitals of Nb atom
and Ising-type SOC are included.
Numerical calculation shows that owing to the broken inversion symmetry, the nonlinear optical Hall 
currents are generated in odd-number-layered NbSe$_2$,
but absent for even-number-layered NbSe$_2$. 
In particular, under irradiating $y$-polarized visible light, nonlinear
spin Hall current appears.
On the other hand, the $x$-polarized visible light generates
nonlinear charge Hall current.
In further, for even-number-layered NbSe$_2$, the spin and charge Hall current can be generated if 
the out-of-plane mirror and inversion symmetries
 are broken by the application of 
electric fields perpendicular to the plane. 
Thus, we provide nonlinear optical selection rule of spin and charge
Hall currents in few-layered NbSe$_2$, which clarifies the even-odd
effect of layer numbers and light polarization dependencies. 
Our results will serve to design potential opt-spintronics devices on the basis of 2D materials.

This paper is organized as follows.
In Sec.~\ref{sec:level2}, we discuss effective model of even- and
odd-number-layered NbSe$_2$ which includes crystal structure and energy band
structure.
In Sec.~\ref{sec:level3}, we numerically calculate nonlinear optical
spin and charge Hall conductivities for alternating current (AC)
fields and find that the nonlinear
optical Hall currents strongly depend on layer numbers and polarization of incident
light.
In Sec.~\ref{sec:level4}, we discuss that when the crystal symmetry is broken by applying
electric fields, nonlinear optical Hall currents can be generated even
in even-number-layered NbSe$_2$. 
In Sec.~\ref{sec:level5}, we summarize our results.
In addition, in Appendix we show details of matrix elements in an
effective Hamiltonian.
Also, we present contour plots of integrands for nonlinear optical spin and charge Hall
conductivities and energy band structures of bilayer
NbSe$_2$ with the application of electric fields.
In Supplementary Material, we provide the details of the 2nd order nonlinear optical
conductivities, temperature effect for
odd-number-layered NbSe$_2$ and nonlinear
optical Hall conductivity of monolayer MoS$_2$ as a reference example of TMDC semiconductor.~\cite{supplementary}

\section{Model}\label{sec:level2}
\begin{figure*}[!t]
  \begin{center}
    \includegraphics[width=1.0\textwidth]{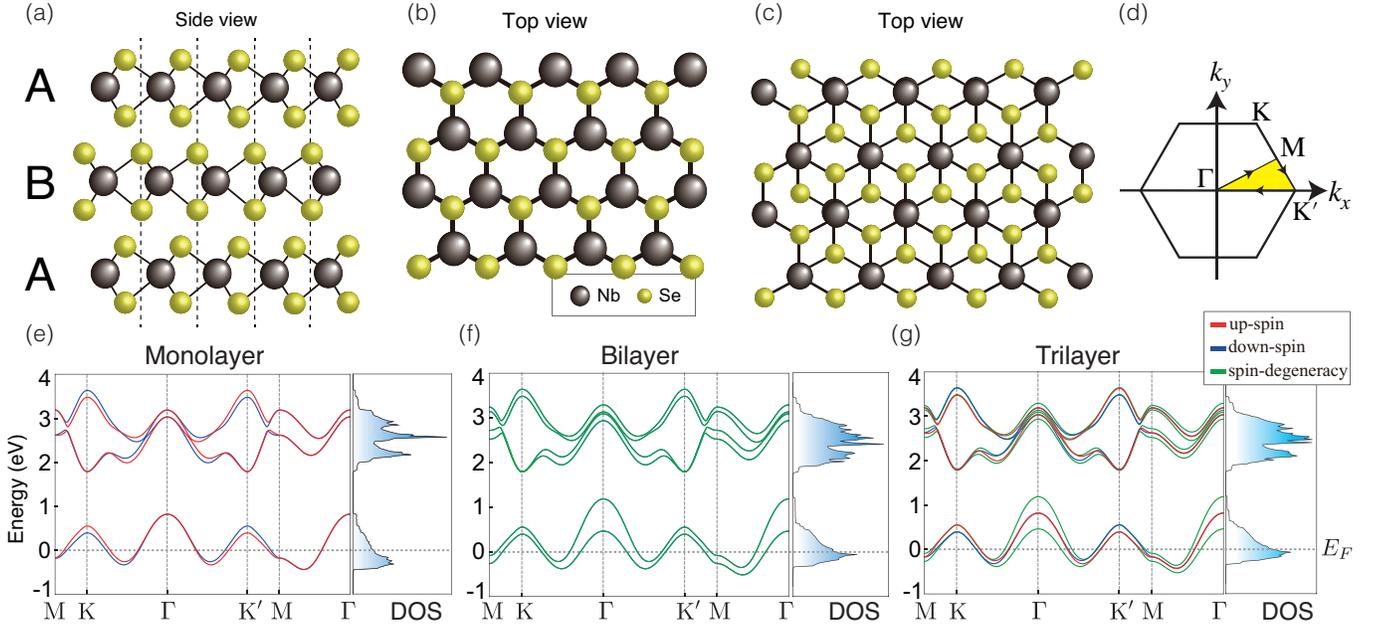}
   \caption{Crystal structures of different few-layered NbSe$_2$ which consist of Nb
   (black) and Se (yellow) atoms. (a) Side view of the lattice structure of monolayer (A), bilayer (AB) and trilayer (ABA) NbSe$_2$.
   Top views of the lattice structures of (b) monolayer, (c) bilayer and trilayer NbSe$_2$.
   (d) 1st BZ of NbSe$_2$. Energy
   band structures and DOS of (e) monolayer, (f) bilayer and (g) trilayer
   NbSe$_2$ with SOC parameter $\lambda_{SOC}=0.0784$ eV, respectively. Fermi level is set to
   zero.}
   \label{fig:1}
  \end{center}
\end{figure*}
In this paper, we show that nonlinear optical spin and charge currents
of few-layered NbSe$_2$ strongly depend on the crystal symmetry and
number of stacking layers.
Especially, we focus on the cases of monolayer, AB-stacked
bilayer and ABA-stacked trilayer NbSe$_2$. 
Figure~\ref{fig:1} (a) shows schematic of few-layered NbSe$_2$ with AB
stacking, which is the most energetically stable stacking sequence in NbSe$_2$. 
Each layer has the out-of-plane mirror symmetry with respect to the plane of
Nb atoms. 
Figures~\ref{fig:1} (b) and (c) show the top views of monolayer NbSe$_2$
and AB-stacked few-layered NbSe$_2$, respectively.
In even-number-layered case, NbSe$_2$ has a space group
D$_{3d}$, which respects to inversion symmetry. However, in case of
odd-number-layered NbSe$_2$, it has a space group D$_{3h}$, which has no spatial
inversion symmetry.
Figure~\ref{fig:1} (d) shows first Brillouin Zone (BZ) for few-layered NbSe$_2$.

We employ a multi-orbitals TBM which includes
$d_{z^2}$, $d_{x^2-y^2}$ and $d_{xy}$ orbitals of Nb atom to describe
the electronic states of NbSe$_2$.~\cite{He, Liu, Ren2021}
The eigenvalue equation for TBM is 
\begin{math}
 \hat{H}(\bm{k})|u_{n\bm{k}}\rangle = E_{n\bm{k}}|u_{n\bm{k}}\rangle,
\end{math}
where $\bm{k}=(k_x,k_y)$ is the wave-number vector, $E_{n\bm{k}}$ is
the eigenvalue and $n=1,2,\cdots,6N$ ($N$ is number of layer) is the band
index.
The eigenvector is defined as 
$|u_{n\bm{k}}\rangle =
(c_{n\bm{k},d_{z^2},\uparrow},c_{n\bm{k},d_{xy},\uparrow},c_{n\bm{k},d_{x^2-y^2},\uparrow},c_{n\bm{k},d_{z^2},\downarrow},c_{n\bm{k},d_{xy},\downarrow},c_{n\bm{k},d_{x^2-y^2},\downarrow})^T$,
where $(\cdots)^T$ indicates the transpose of vector and
$c_{n\bm{k}\tau s}$ means the amplitude at atomic orbital $\tau$ with
spin $s$ for the $n$th energy band at $\bm{k}$.
The Hamiltonian of monolayer NbSe$_2$ with the SOC can be written as
\begin{equation}
 \hat{H}_{mono}(\bm{k})=\hat{\sigma}_0\otimes \hat{H}_{TNN}(\bm{k})+\hat{\sigma}_z\otimes\frac{1}{2}\lambda_{SOC} \hat{L}_z
\end{equation}
with
\begin{equation}
 \hat{H}_{TNN}(\bm{k})=
  \begin{pmatrix}
   V_{0}&V_{1}&V_{2}\\
   V_{1}^{*}&V_{11}&V_{12}\\
   V_{2}^{*}&V_{12}^{*}&V_{22}\\
  \end{pmatrix}
\end{equation}
and
\begin{equation}
 \hat{L}_z=
  \begin{pmatrix}
   0&0&0\\
   0&0&-2i\\
   0&2i&0\\
  \end{pmatrix}.
\end{equation}
Here, $\hat{\sigma}_0$ and $\hat{\sigma}_z$ are Pauli matrices and
$\lambda_{SOC}$ is the Ising-type SOC parameter. In monolayer
NbSe$_2$,
$\lambda_{SOC}=0.0784$ eV.
$\hat{H}_{TNN}(\bm{k})$ includes the electron hoppings only among three
$d$-orbitals of Nb atoms, which are assumed up to third-nearest neighbor
sites as shown in Appendix~A.
Similarly, Hamiltonians of bilayer and trilayer NbSe$_2$ can be obtained as
\begin{equation}
 \hat{H}_{bi}(\bm{k})=
  \begin{pmatrix}
   \hat{H}_{mono}(-\bm{k})&\hat{H}_{int}(\bm{k})\\
   \hat{H}^{\dag}_{int}(\bm{k})&\hat{H}_{mono}(\bm{k})\\
  \end{pmatrix}
\end{equation}
and
\begin{equation}
 \hat{H}_{tri}(\bm{k})=
  \begin{pmatrix}
   \hat{H}_{mono}(\bm{k})&\hat{H}_{int}(\bm{k})&0\\
   \hat{H}^{\dag}_{int}(\bm{k})&\hat{H}_{mono}(-\bm{k})&\hat{H}_{int}(\bm{k})\\
   0&\hat{H}^{\dag}_{int}(\bm{k})&\hat{H}_{mono}(\bm{k})\\
  \end{pmatrix},
\end{equation}
respectively.~\cite{Sohn2018}
Here, interlayer coupling Hamiltonian $\hat{H}_{int}(\bm{k})$ is
considered as
\begin{equation}
 \hat{H}_{int}(\bm{k})=
  \begin{pmatrix}
   T_{01}&0&0\\
   0&T_{02}&0\\
   0&0&T_{02}\\
  \end{pmatrix}.
\end{equation}
The details of matrix elements $V_0$, $V_1$, $V_2$, $V_{11}$,
$V_{12}$, $V_{22}$, $T_{01}$ and $T_{02}$
can be found in Appendix~A.

Figures~\ref{fig:1} (e), (f) and (g) show 
the energy band structures of monolayer, bilayer and trilayer NbSe$_2$ 
together with the corresponding density of
states (DOS), respectively. 
Here, red, blue and green lines indicate spin-up, spin-down and spin-degenerated states.
NbSe$_2$ is metallic, but a large energy band gap between
the partially filled valence bands and empty conduction bands.
Also, opposite spin splitting in the energy band structure can be seen at the valence
band edges in K and K$^{\prime}$ points in monolayer 
NbSe$_2$ owing to the broken inversion symmetry.
However, because even-number-layered NbSe$_2$ such as bilayer respect the inversion
symmetry, it does not show the spin splitting.
It should be noted that even-number-layered NbSe$_2$ has larger band
splitting at valence band in $\Gamma$ point, becase the interlayer
interaction becomes larger in $\Gamma$ point than in K and K$^\prime$ points.
Figure~\ref{fig:1} (g) shows the calculated energy band structure of trilayer
NbSe$_2$,
which can be understood by overwriting spin degenerated energy band structure of bilayer
NbSe$_2$ onto that of spin-splitting energy dispersion of monolayer
NbSe$_2$. 
Since the inversion symmetry is broken in odd-number-layered NbSe$_2$,
spin degeneracy is lifted. 
However, owing to the existence of spin degenerated energy band of bilayer
NbSe$_2$, the spin splitting at K and K$^\prime$ points is not clearly
seen. The details of Fermi spin-dependent surface structures are shown
in Appendix A.

\section{Nonlinear optical Hall conductivity}\label{sec:level3}
\begin{table*}
\caption{\label{tab:table1}Nonlinear optical spin and charge conductivities of mono, bi and
  trilayer NbSe$_2$. The conductivities of monolayer NbSe$_2$ are summarized in Figs.~S2.}
\begin{ruledtabular}
\begin{tabular}{ccc}
 &nonlinear optical spin conductivity&nonlinear optical charge conductivity \\ \hline
  mono (D$_{3h}$)&$\sigma^{\rm{spin}}_{xxx}(\omega, \omega)=-\sigma^{\rm{spin}}_{xyy}(\omega, \omega)$&$\sigma^{\rm{charge}}_{yyy}(\omega, \omega)=-\sigma^{\rm{charge}}_{yxx}(\omega, \omega)$ \\
  &$=-\sigma^{\rm{spin}}_{yxy}(\omega, \omega)=-\sigma^{\rm{spin}}_{yyx}(\omega, \omega)$&$=-\sigma^{\rm{charge}}_{xyx}(\omega, \omega)=-\sigma^{\rm{charge}}_{xxy}(\omega, \omega)$ \\
 bi (D$_{3d}$)&zero&zero \\
 tri (D$_{3h}$)&$\sigma^{\rm{spin}}_{xxx}(\omega, \omega)=-\sigma^{\rm{spin}}_{xyy}(\omega, \omega)$&$\sigma^{\rm{charge}}_{yyy}(\omega, \omega)=-\sigma^{\rm{charge}}_{yxx}(\omega, \omega)$ \\
 &$=-\sigma^{\rm{spin}}_{yxy}(\omega, \omega)=-\sigma^{\rm{spin}}_{yyx}(\omega, \omega)$&$=-\sigma^{\rm{charge}}_{xyx}(\omega, \omega)=-\sigma^{\rm{charge}}_{xxy}(\omega, \omega)$ \\
\end{tabular}
\end{ruledtabular}
\end{table*}
\begin{figure*}[t]
  \begin{center}
    \includegraphics[width=0.95\textwidth]{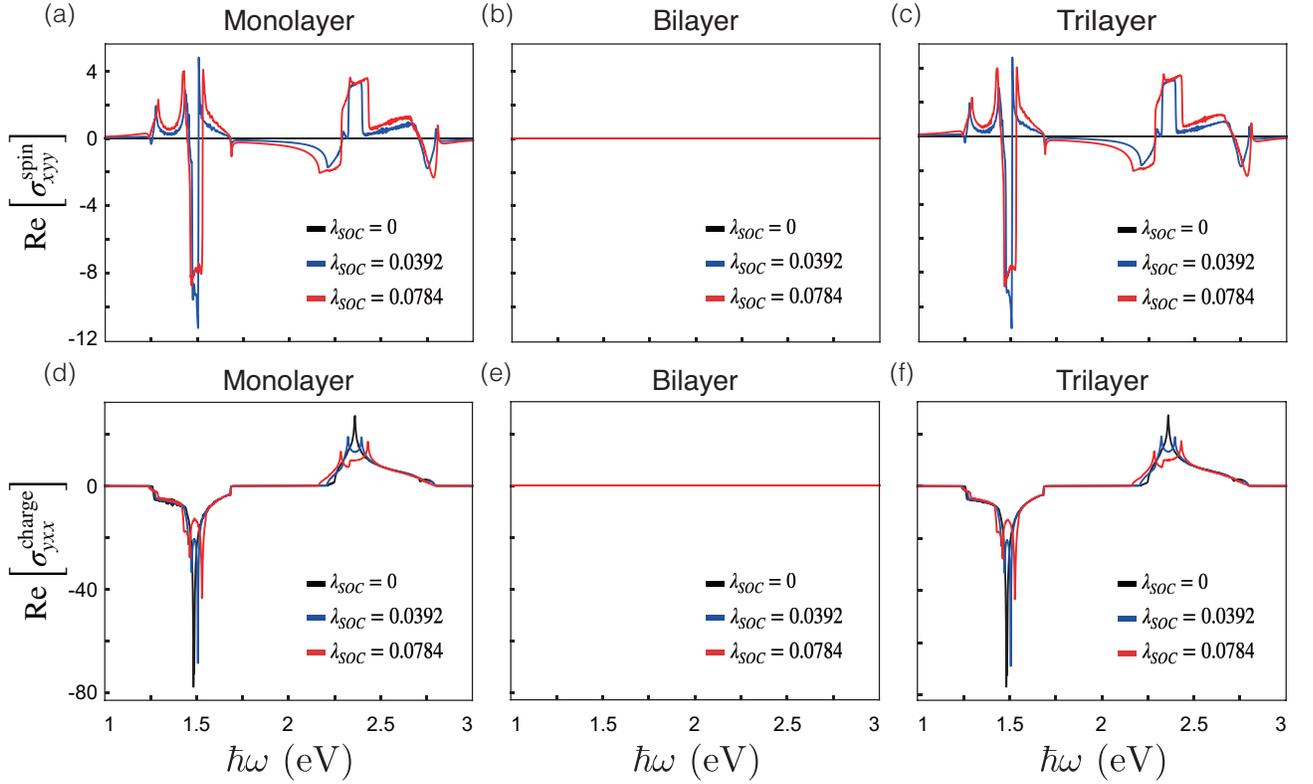}
    \caption{Real parts of nonlinear optical spin Hall conductivities
      $\mathrm{Re}\left[\sigma^{\rm{spin}}_{xyy}(\omega, \omega)\right]$ of (a) monolayer, (b) bilayer
      and (c) trilayer NbSe$_2$, respectively.
      By irradiating $y$-polarized light, nonlinear
      optical spin Hall current is generated in odd-number-layered
      NbSe$_2$.
      Real parts of nonlinear optical charge Hall
      conductivities $\mathrm{Re}\left[\sigma^{\rm{charge}}_{yxx}(\omega, \omega)\right]$ of (d) monolayer, (e)
      bilayer and (f) trilayer NbSe$_2$, respectively.
      Nonlinear optical charge Hall current is generated by
      $x$-polarized light irradiation. Red, blue and black lines
      indicate several different SOC parameters. The units of nonlinear optical spin and charge Hall conductivities are $e^2$ and $e^3/\hbar$, respectively.}
    \label{fig:2}
  \end{center}
\end{figure*}
We numerically calculate nonlinear optical spin and charge Hall conductivities for few-layered NbSe$_2$ based on an effective TBM.
In general, the 2nd order nonlinear optical spin conductivity can be given as
~\cite{Kang2010, Lee2002, Kang2012, Kang2013, Moss1987,
  Ghahramani1991, Rashkeev2001, Leitsmann2005, Aversa1995, Sipe1993PRB} 
\begin{equation}
    \sigma^{\rm{spin}}_{ijk}(\omega_1, \omega_2) \equiv -\frac{\hbar^2e^2}{S}\sum_{\bm{k}}\Omega^{\rm{spin}}_{ijk} (\omega_1, \omega_2, \bm{k})
\label{eq:nonform}
\end{equation}
with
\begin{equation}
 \begin{split}
  \Omega^{\rm{spin}}_{ijk} & (\omega_1, \omega_2, \bm{k})=
  \sum_{nml}\frac{1}{E_{ml}E_{ln}(E_{mn}-\hbar\omega_1-\hbar\omega_2-i\eta)} \\
& \times\Biggl[\frac{\braket{u_{n\bm{k}}|\hat{v}_j|u_{l\bm{k}}}\braket{u_{l\bm{k}}|\hat{v}_k|u_{m\bm{k}}}\braket{u_{m\bm{k}}|\hat{j}^{\rm{spin}}_i|u_{n\bm{k}}} f_{ml}}{E_{ml}-\hbar\omega_2-i\eta} \\
& -\frac{\braket{u_{n\bm{k}}|\hat{v}_k|u_{l\bm{k}}}\braket{u_{l\bm{k}}|\hat{v}_j|u_{m\bm{k}}}\braket{u_{m\bm{k}}|\hat{j}^{\rm{spin}}_i|u_{n\bm{k}}} f_{ln}}{E_{ln}-\hbar\omega_2-i\eta}\Biggl],
 \end{split}
\label{eq:nonformo}
\end{equation}
where $\Omega^{\rm{spin}}_{ijk}(\omega_1, \omega_2, \bm{k})$ is
integrand of nonlinear optical spin conductivity. Here, $i$($j$, $k$)
indicates the direction $x$ or $y$.
In particular, $i$ is the generation direction of nonlinear optical
spin current, and $j$($k$) is polarization of incident light.
Also, $n$($m$, $l$) is the band index including spin degree of freedom, $\Ket{u_{n\bm{k}}}$ is the eigen
function with the eigen energy $E_{n\bm{k}}$ and $f(E_{n\bm{k}})$ is Fermi-Dirac
distribution function.
$E_{ml}\equiv E_{m}-E_{l}$, $f_{ml}\equiv
f(E_{m\bm{k}})-f(E_{l\bm{k}})$ and $\omega_q$ means the $q$th
frequency mode.
$\eta$ is infinitesimally small real number and $S$ is the area of system.
Moreover, $\hat{j}^{\rm{spin}}_i$ is the spin current operator and written as
$\hat{j}^{\rm{spin}}_i=\frac{1}{2}\{\frac{\hbar}{2}\hat{\sigma}_z\otimes \hat{I}_N, \hat{v}_i\}$,
where $\hat{I}_N$ is the $N\times N$ identity matrix and $N=3,6,9$ is
used for monolayer, bilayer and trilayer NbSe$_2$, respectively.
Here, $\hat{v}_{i}=\frac{1}{\hbar}\frac{\partial \hat{H}}{\partial i}$ is the group velocity operator.
The nonlinear optical charge conductivity
$\sigma^{\rm{charge}}_{ijk}(\omega_1, \omega_2)$ is also obtained by
changing $\hat{j}^{\rm{spin}}_i$ to $\hat{j}^{\rm{charge}}_i$ which is
defined as
$\hat{j}^{\rm{charge}}_i=\frac{1}{2}\{-e\hat{\sigma}_0\otimes
\hat{I}_N, \hat{v}_i\}$.
We add the superscript ``$\rm{spin}$'' for the nonlinear optical spin conductivity in order to distinguish its
conductivity from the nonlinear optical charge conductivity.
In case of $\omega_1+\omega_2 =\omega_3$, the process is called as SFG [see Figs.~S3].~\cite{supplementary}
Especially, in case of $\omega_1=\omega_2\equiv \omega$, the process
is called as SHG.
Also, in case of $\omega_1-\omega_2=\omega_3$, the process is called
as difference frequency generation (DFG).~\cite{Lu2011, Axel2012}
Since we have interest in SHG, we focus on the SHG process in this
manuscript, i.e. $\omega_1=\omega_2= \omega$.

The 2nd order nonlinear optical conductivity expressed by Eq.~(\ref{eq:nonform})
can be separated into two inter-band processes: (i) optical transition between two
bands $\sigma^{(2)}_{ijk}$ and (ii) optical transition involving three
bands $\sigma^{(3)}_{ijk}$.~\cite{Aversa1995, Passos2021} Namely, $\sigma_{ijk}$
can be decomposed into 
\begin{equation}
  \sigma_{ijk}(\omega_1, \omega_2)=\sigma^{(2)}_{ijk}(\omega_1, \omega_2)+\sigma^{(3)}_{ijk}(\omega_1, \omega_2).
\label{eq:2and3bands}
\end{equation}
The detail of derivation can be found in Appendix B.
In particular, considering the inter-band transition between two bands around Fermi surface, $\sigma^{(2)}_{ijk}$ can be rewritten as
\begin{equation}
    \sigma^{(2)}_{ijk}(\omega_1, \omega_2\rightarrow 0)=-\frac{i
     e^2}{S}\frac{1}{\hbar\omega_1+i\eta}D_i,
\label{eq:2bands}
\end{equation}
where $D_i$ is the Berry curavture dipole~\cite{Zeng2021,Sodemann2015} 
\begin{equation}
D_i=\sum_{\bm{k}}\sum_{n}\Omega_{n}(\bm{k})\braket{u_{n\bm{k}}|\hat{j}_i|u_{n\bm{k}}}
 \frac{\partial f_{n\bm{k}}}{\partial E_{n\bm{k}}}. 
\end{equation}
Thus, 
the two-band process is nothing more than the effect of Berry curvature dipole.
In the direct current (DC) limit, i.e.,
$\omega_1\rightarrow 0$, $D_i$ is identically
zero in NbSe$_2$. However, it is expected to be finite even in the DC limit, if the uniaxial
strain is applied to the system.~\cite{Zeng2021,Sodemann2015} 
In this work, we consider $\sigma_{ijk}$ regardless of two-bands and
three-bands inter-band processes.

Since the nonlinear optical conductivity $\sigma_{ijk}$ is the $3$rd
rank tensor, in general $\sigma^{\rm{spin}}_{ijk}$ and $\sigma^{\rm{charge}}_{ijk}$
have the $3^3=27$ components, respectively.
However, using the Neumann's principle,~\cite{Wen2019, Lucas2021}
we can find the nonvanishing elements of $\sigma^{\rm{spin}}_{ijk}$ and
$\sigma^{\rm{charge}}_{ijk}$
from the crystal symmetry. 
Because the inversion symmetry is broken in monolayer NbSe$_2$, 
the nonvanishing tensor elements are obtained as following:
\begin{equation}
  \begin{split}
    &\sigma^{\rm{spin}}_{xxx}(\omega, \omega)=-\sigma^{\rm{spin}}_{xyy}(\omega, \omega) \\
    &=-\sigma^{\rm{spin}}_{yxy}(\omega, \omega)=-\sigma^{\rm{spin}}_{yyx}(\omega, \omega)
   \end{split}
\label{eq.spinnoum}
\end{equation}
and
\begin{equation}
  \begin{split}
    &\sigma^{\rm{charge}}_{yyy}(\omega, \omega)=-\sigma^{\rm{charge}}_{yxx}(\omega, \omega) \\
    &=-\sigma^{\rm{charge}}_{xyx}(\omega, \omega)=-\sigma^{\rm{charge}}_{xxy}(\omega, \omega).
   \end{split}
\label{eq.chargenoum}
\end{equation}
However, owing to the inversion symmetry in even-number-layered
NbSe$_2$ such as bilayer NbSe$_2$, the nonlinear optical conductivities
are obviously absent, i.e., all the tensor elements are identically zero.
Also, since the crystal symmetry of trilayer NbSe$_2$ is 
identical to the monolayer NbSe$_2$, same relations of 
the nonlinear optical spin and charge conductivities are obtained using Neumann's principle.
Table~\ref{tab:table1} summarizes the relations of nonvanishing tensor
elements of $\sigma^{\rm{spin}}_{ijk}$ and $\sigma^{\rm{charge}}_{ijk}$
for few-layered NbSe$_2$. 
It should be noted that 
the nonlinear optical charge conductivity is zero,
whenever the nonlinear optical spin conductivity has finite value.
Since the 2nd order nonlinear optical conductivity $\sigma_{ijk}$
is a complex function of $\omega$, the conductivity can be separated as
\begin{equation}
  \sigma_{ijk}(\omega,
   \omega)=\mathrm{Re}\left[\sigma_{ijk}(\omega, \omega)\right]
+i\mathrm{Im}\left[\sigma_{ijk}(\omega, \omega)\right],
\end{equation}
where $\mathrm{Re}\left[\sigma_{ijk}(\omega, \omega)\right]$ and 
$\mathrm{Im}\left[\sigma_{ijk}(\omega, \omega)\right]$ are real and imaginary
parts, respectively. 
In the main text, we focus on the real part of $\sigma_{ijk}$.
The details of the imaginary part of $\sigma_{ijk}$ are shown in Supplementary Material.

Figures~\ref{fig:2} (a), (b) and (c) show the real parts of nonlinear optical spin
Hall conductivities $\mathrm{Re}\left[\sigma^{\rm{spin}}_{xyy} (\omega, \omega)\right]$ of monolayer, bilayer and trilayer NbSe$_2$,
respectively.
Here, $\mathrm{Re}\left[\sigma^{\rm{spin}}_{xyy} (\omega, \omega)\right]$ is
considered as the case of SHG process and has Ising-type SOC parameter
$\lambda_{SOC}=0.0784$ eV.
Also, the cases for $\lambda_{SOC}=0.0392$ and $0$ eV are plotted for
the comparison.
$\mathrm{Re}\left[\sigma^{\rm{spin}}_{xyy} (\omega, \omega)\right]$ represents that the spin Hall current
is generated in $x$ direction by irradiating $y$ polarized light.
It mainly has two peaks around $1.5$ and $2.5$
eV for odd-number-layered NbSe$_2$ owing to even-parity with respect
to $k_x$ and $k_y$ axes in contour plot of $\Omega^{\rm{spin}}_{xyy} (\omega, \omega, \bm{k})$ [see Appendix~C].
One peak can be seen around $1.5$ eV, and indicates an excitation from
valence band to conduction band by one incident photon at $2\hbar\omega$.
The other peak around $2.5$ eV shows that two incident photons at $\hbar\omega$
occurs an excitation from valence band to intermediate band and then conduction band.

Figures~\ref{fig:2} (d), (e) and (f) show the real parts of nonlinear optical charge
Hall conductivities $\mathrm{Re}\left[\sigma^{\rm{charge}}_{yxx} (\omega, \omega)\right]$ of SHG
process for monolayer, bilayer and trilayer NbSe$_2$, respectively.
$\mathrm{Re}\left[\sigma^{\rm{charge}}_{yxx} (\omega, \omega)\right]$ represents that the charge Hall current is generated in $y$ direction by
irradiating $x$ polarized light.
There are mainly two peaks around $1.5$ and $2.5$ eV as same as the
case of $\mathrm{Re}\left[\sigma^{\rm{spin}}_{xyy} (\omega, \omega)\right]$, which appear for
$\mathrm{Re}\left[\sigma^{\rm{charge}}_{yxx} (\omega, \omega)\right]$ owing to the asymmetry of $\Omega^{\rm{charge}}_{yxx} (\omega, \omega, \bm{k})$ with
respect to $k_x$ axis in contour plot [see Appendix~C].
For one incident photon at $2\hbar\omega$, one peak appears around $1.5$
eV, and it is larger than the other peak around $2.5$ eV which can be seen
by two incident photons at $\hbar\omega$.

Thus, we can find that the nonlinear optical spin and charge Hall
currents strongly depend on layer numbers and polarization of incident
light in visible range, i.e., nonlinear optical selection rule of spin
and charge Hall currents in few-layered NbSe$_2$.

Here we briefly mention the extreme similarity of 
the nonlinear optical conductivities 
between monolayer and trilayer NbSe$_2$ as shown in Fig.~\ref{fig:2}. 
The trilayer NbSe$_2$ can be viewed as the composite of monolayer and
bilayer NbSe$_2$. Therefore, we have three contributions of optical
transition processes: (A) intralayer optical transition of monolayer
NbSe$_2$, (B) intra- and interlayer optical transition of bilayer
NbSe$_2$, and (C) interlayer optical transitions between monolayer and
bilayer NbSe$_2$. 
Since the process (B) is identically zero, the process (A) dominates the
optical conductivities of trilayer NbSe$_2$. It is shown that 
the process (C) is canceled because of band inversion along the 
$\Gamma$-K and $\Gamma$-K$^{\prime}$ lines in BZ. 
The detail can be found in Supplementary Material.

Also, we have mentioned that the nonlinear optical Hall
conductivity has the peaks (around $1.5$ and $2.5$ eV) corresponding to absorption of two
photons in odd-number-layered NbSe$_2$, which cannot be seen for
linearly optical Hall conductivity obtained by Kubo
formula.~\cite{Ren2021, Kim2021}
It should be noted that the magnitude of peak clearly corresponds to DOS of
NbSe$_2$.
The details of these peaks are shown in Supplementary Material.

In addition, TMDC semiconductor such as MoS$_2$ has a pronounced peak
around $1.75$ eV in the nonlinear optical Hall conductivity.
The details of nonlinear optical Hall conductivity of monolayer MoS$_2$
can be found in Supplementary Material. 
This transition process corresponds to the SHG process marked with the green arrows 
in Fig.~\ref{fig:5} (c) of Appendix~A.
Thus, SHG can be expected in the doped MoS$_2$. 
Similarly, when we consider the case of 
electron-doped NbSe$_2$ to make the valence bands fully occupied,
the system behaves as a semiconductor. In this case, 
the nonlinear optical conductivity for SHG process has the localized
peak around $1.5$ eV (not shown).
This transition process also corresponds to 
the SHG process marked with the green arrows 
in Fig.~\ref{fig:5} (c) of Appendix~A, which is same as the
case of MoS$_2$.

In previous works,~\cite{Xiao2020, Zhou2020, Ma2019, Sodemann2015} 
it is reported that 
the nonlinear charge Hall current in DC limit 
can be
generated
even in the absence of magnetic field, by considering monolayer NbSe$_2$ under uniaxial strain
or hole-doped semiconductor TMDC. 
These results can be induced by Berry curvature dipole which provides
unconventional behavior, i.e., nonlinear Hall effect, but it is
limited to in metallic TMDC.
In this work, however, we can show that 
the nonlinear charge Hall current appears in not only metallic TMDC
but also semiconductor TMDC by light irradiation even without SOC
and uniaxial strain [see Figs.~\ref{fig:2} (d) and (f)]. 
Thus, we can generate the nonlinear charge Hall current simply
by irradiating light in few-layered NbSe$_2$ without the 
external perturbations such as magnetic field, strain and carrier dopings.

\begin{figure}[t]
  \begin{center}
    \includegraphics[width=0.5\textwidth]{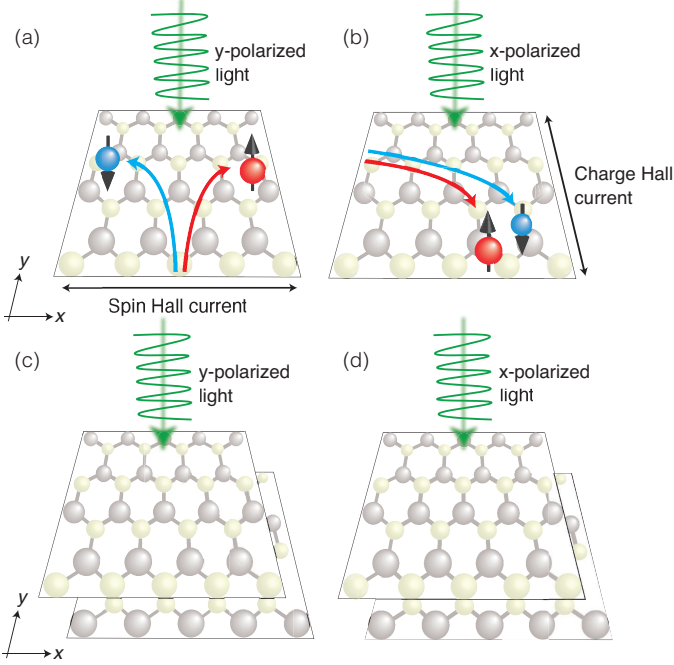}
    \caption{Schematics of nonlinear optical spin and charge Hall currents in few-layered NbSe$_2$.
    (a) By $y$-polarized light irradiation, nonlinear optical spin
      Hall current is generated in $x$ direction, but nonlinear
      optical charge Hall current is absent in odd-number-layered NbSe$_2$.
    (b) By $x$-polarized light irradiation, nonlinear optical spin
      Hall current disappears, but nonlinear optical charge Hall
      current is generated in $y$ direction in odd-number-layered
      NbSe$_2$.
    Nonlinear optical (c) spin and (d) charge Hall currents are not generated in even-number-layered NbSe$_2$.}
    \label{fig:3}
  \end{center}
\end{figure}
Figure~\ref{fig:3} summarizes the schematics of nonlinear optical Hall
currents in few-layered NbSe$_2$ which capture the results of
Fig.~\ref{fig:2}.
Figure~\ref{fig:3} (a) indicates that nonlinear optical spin Hall
current is generated in $x$ direction by irradiating $y$ polarized
light in odd-number-layered NbSe$_2$.
However, nonlinear optical charge Hall current is absent. 
Figure~\ref{fig:3} (b) shows that nonlinear optical charge Hall
current is generated in $y$ direction by irradiating $x$ polarized
light, but then the spin Hall current is absent.
Also, Figures~\ref{fig:3} (c) and (d) show that because
even-number-layered NbSe$_2$ respects to inversion symmetry, the
nonlinear optical spin and charge 
Hall currents are identically zero.

\section{Electric field effect of nonlinear optical Hall conductivity}\label{sec:level4}
\begin{figure}[h]
  \begin{center}
    \includegraphics[clip, width=0.374\textwidth]{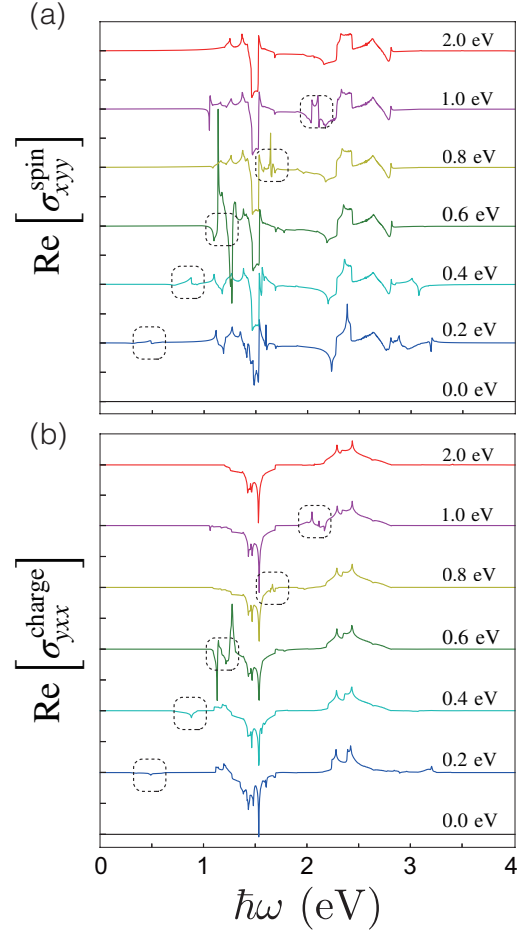}
    \caption{Nonlinear optical (a) spin and (b) charge Hall
   conductivities of bilayer NbSe$_2$ for several different applied electric fields.
      The electric fields are $F=0.2$ (blue), $0.4$
      (cyan), $0.6$ (green), $0.8$ (yellow), $1.0$ (purple) and $2.0$ eV (red),
      respectively.
      (a) By irradiating $y$-polarized light, real part of nonlinear optical spin
      Hall current $\mathrm{Re}\left[\sigma^{\rm{spin}}_{xyy} (\omega, \omega)\right]$ is generated in
      bilayer NbSe$_2$ with the application of electric fields.
      (b) By irradiating $x$-polarized light, real part of nonlinear optical charge
      Hall current $\mathrm{Re}\left[\sigma^{\rm{charge}}_{yxx} (\omega, \omega)\right]$ is generated
      in bilayer NbSe$_2$. The units of nonlinear optical spin and charge Hall
   conductivities are $e^2$ and $e^3/\hbar$, respectively.} 
    \label{fig:4}
  \end{center}
\end{figure}

Since the application of electric field perpendicular to the plane
breaks the crystal inversion symmetry in bilayer NbSe$_2$, 
the nonlinear optical spin and charge Hall conductivities can be
generated even in bilayer NbSe$_2$ with the application of electric field.
Figure~\ref{fig:4} (a) shows the real part of nonlinear optical spin Hall
conductivity $\mathrm{Re}\left[\sigma^{\rm{spin}}_{xyy} (\omega, \omega)\right]$ of
bilayer NbSe$_2$ for several different applied electric
fields.
Here, black, blue, cyan, green, yellow, purple and red lines indicate
the applied electric fields: $F=0.0$, $0.2$, $0.4$,
$0.6$, $0.8$, $1.0$ and $2.0$ eV, respectively.
We can see peaks of $\mathrm{Re}\left[\sigma^{\rm{spin}}_{xyy} (\omega, \omega)\right]$ around $1.5$
and $2.5$ eV as same as the case of odd-number-layer.
In addition, there is a pronounced peak which shifts toward higher
frequency with increase of electric field, which is indicated by the
dashed square in Fig.~\ref{fig:4} (a).
The peak is originated from the interlayer optical absorption. 
Since the energy bands of upper (lower) layer shift toward higher
(lower) energy, the energy difference between upper and lower layers
increases with increase of electric field, resulting in the shift of interlayer optical absorption peak.
The details about energy band structures of bilayer NbSe$_2$ with applied
electric fields and its parity between layers are shown in Appendix~D.

Figure~\ref{fig:4} (b) shows the real part of nonlinear optical charge Hall
conductivity $\mathrm{Re}\left[\sigma^{\rm{charge}}_{yxx} (\omega, \omega)\right]$ of
bilayer NbSe$_2$ with applied electric fields.
Because of the broken crystal inversion symmetry in even-number-layered
NbSe$_2$ with applied electric fields, the nonlinear optical charge Hall
current can be generated by irradiating $x$ 
polarized light.
$\mathrm{Re}\left[\sigma^{\rm{charge}}_{yxx} (\omega, \omega)\right]$ also has peaks around $1.5$ and $2.5$ eV,
which is similar to the case of $\mathrm{Re}\left[\sigma^{\rm{spin}}_{xyy} (\omega, \omega)\right]$.
Similarly, we can observe the frequency shift of interlayer optical
absorption peak which is indicated by the dashed squares. 
Thus, we can indicate that owing to the broken inversion symmetry in
even-number-layered NbSe$_2$ with applied electric fields, the
nonlinear optical spin and charge Hall currents can be generated by irradiating visible
light.

Instead of the application of electric fields to NbSe$_2$, we consider the nonlinear optical spin and charge
conductivities of bilayer NbSe$_2$ with each layer having a different
Fermi energy, i.e. decoupled bilayer NbSe$_2$.
The details of the nonlinear optical spin and charge
Hall conductivities of the decoupled bilayer NbSe$_2$ are shown in Appendix~E.

\section{Conclusion}\label{sec:level5}
In conclusion, we have theoretically proposed that nonlinear optical
spin and charge Hall currents based on SHG process can be enhanced by irradiating visible light.
Also, we have shown that the Hall currents strongly depend on layer
numbers, crystal symmetry of NbSe$_2$ and polarization of incident
light, i.e., nonlinear optical selection rule of spin and charge Hall
currents in few-layered NbSe$_2$. 
In previous works, it is known that the nonlinear Hall effect is
induced in metallic and doped semiconductor TMDCs by uniaxial strain,
which depends on the Berry curvature dipole in 1st BZ.
In our work, we can find nonlinear Hall effect by light irradiation
in NbSe$_2$ and MoS$_2$ with hole-doping, even in the absence of strain.
In Supplementaly Material, 
it is also shown that the nonlinear optical spin and charge Hall
conductivities can occur in MoS$_2$ without doping.
Thus, in general, it is expected that the nonlinear optical Hall effect can occur in TMDC materials.

In addition, we have found that the nonlinear optical spin and charge
Hall currents of few-layered NbSe$_2$ based on the effective TBM are
robust to temperature and are expected to be observed even at room
temperature [see Supplementary Material].

In this paper, we have found that the oscillating spin and charge Hall current
could be induced by the SHG. Though the static charge and spin
accumulation does not occur, the polarized spin current can be extracted
if we attach the half-metal materials to the
edge of the sample as a spin filter.

Thus, few-layered NbSe2 can be used for the source of induced
nonlinear optical spin Hall current by SHG. 
Our results can serve to design opt-spintronics devices on
the basis of 2D materials.

\begin{acknowledgments}
This work was supported by JSPS KAKENHI
(Nos. JP21H01019, JP18H01154) and JST CREST (No. JPMJCR19T1).
\end{acknowledgments}

\appendix

\section{Matrix elements of few-layered NbSe$_2$}
\begin{table*}[t]
\caption{\label{tab:table2}Fitting parameters for the effective TBM Hamiltonian of
 few-layered NbSe$_2$. The energy parameters $\epsilon_1$ to
 $\lambda_{SOC}$ are in units of eV.}
\begin{ruledtabular}
\begin{tabular}{ccccccccccc}
$\epsilon_1$&$\epsilon_2$&$t_0$&$t_1$&$t_2$&
$t_{11}$&$t_{12}$&$t_{22}$&$r_0$&$r_1$&$r_2$\\
$r_{11}$&$r_{12}$&$u_0$&$u_1$&
$u_2$&$u_{11}$&$u_{12}$&$u_{22}$&$t_{01}$&$t_{02}$&$\lambda_{SOC}$\\
\hline
1.4466 & 1.8496 & -0.2308 & 0.3116 & 0.3459 & 0.2795 & 0.2787 & -0.0539
			     & 0.0037 & -0.0997 & 0.0385 \\
0.0320 & 0.0986 & 0.1233 & -0.0381 & 0.0535 & 0.0601 & -0.0179
			     & -0.0425 & -0.0179 & -0.0702 & 0.0784 \\
\end{tabular}
\end{ruledtabular}
\end{table*}
\begin{figure*}[t]
  \begin{center}
    \includegraphics[width=0.91\textwidth]{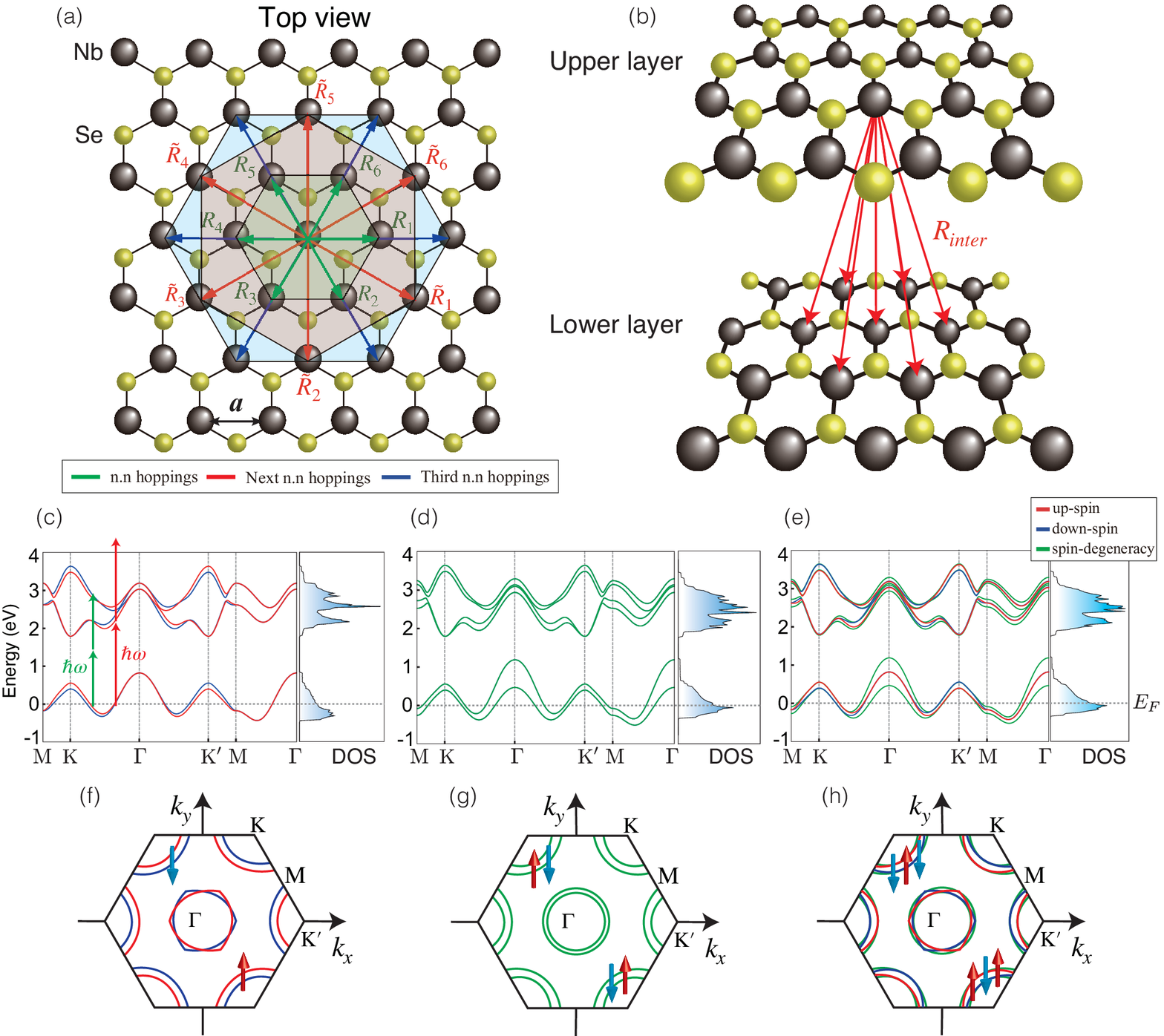}
   \caption{(a) Top view of crystal structure of monolayer NbSe$_2$
   which consists of Nb (black) and Se (yellow) atoms. Green, red and
   blue arrows indicate hopping vectors
   $\bm{R_i}$ $(i=1,2,\cdots,6)$ pointing to n.n sites,
   the vectors $\bm{\tilde{R_j}}$ $(j=1,2,\cdots,6)$ pointing to next n.n
   sites and the vectors $2\bm{R_i}$ pointing to third n.n sites, respectively. $a$ is the
   lattice constant. (b) Side view of crystal structure of bilayer NbSe$_2$.
   Interlayer hopping vector $\bm{R}_{inter}$ points $d$-orbital of upper layer to n.n sites of lower layer.
   Energy band structures and DOS of (c) monolayer, (d) bilayer and
   (e) trilayer NbSe$_2$ with SOC parameter $\lambda_{SOC}=0.0784$ eV, respectively.
   Fermi level is set to zero.
The arrows in (c) include the two SHG processes: (i) inter-band transition by two-photon absorption (green arrows) and (ii) inter-band transiton even by one-photon absorption (red arrows).
   Fermi surface of (f) monolayer, (g) bilayer and (h) trilayer
   NbSe$_2$, where red, blue and green lines indicate for up-spin, down-spin and
   spin-degenerated states, respectively. 
}
    \label{fig:5}
  \end{center}
\end{figure*}
We employ a multi-orbitals TBM which includes $d_{z^2}$, $d_{x^2-y^2}$ and $d_{xy}$ orbitals of Nb atom to describe the electronic states of NbSe$_2$.
The eigenvalue equation for TBM is 
\begin{math}
 \hat{H}(\bm{k})|u_{n\bm{k}}\rangle = E_{n\bm{k}}|u_{n\bm{k}}\rangle,
\end{math}
where $\bm{k}=(k_x,k_y)$ is the wave-number vector, $E_{n\bm{k}}$ is
the eigenvalue and $n=1,2,\cdots,6N$ ($N$ is number of layers) is the band index.
The eigenvector is defined as 
$|u_{n\bm{k}}\rangle =
(c_{n\bm{k},d_{z^2},\uparrow},c_{n\bm{k},d_{xy},\uparrow},c_{n\bm{k},d_{x^2-y^2},\uparrow},c_{n\bm{k},d_{z^2},\downarrow},c_{n\bm{k},d_{xy},\downarrow},c_{n\bm{k},d_{x^2-y^2},\downarrow})^T$, 
where $(\cdots)^T$ indicates the transpose of vector and $c_{n\bm{k}\tau s}$ means the amplitude at atomic orbital $\tau$ with spin $s$ for the $n$th energy band at $\bm{k}$.
The Hamiltonian of monolayer NbSe$_2$ with Ising-type SOC can be written as
\begin{equation}
 \hat{H}_{mono}(\bm{k})=\hat{\sigma}_0\otimes \hat{H}_{TNN}(\bm{k})+\hat{\sigma}_z\otimes\frac{1}{2}\lambda_{SOC} \hat{L}_z
\end{equation}
with
\begin{equation}
 \hat{H}_{TNN}(\bm{k})=
  \begin{pmatrix}
   V_{0}&V_{1}&V_{2}\\
   V_{1}^{*}&V_{11}&V_{12}\\
   V_{2}^{*}&V_{12}^{*}&V_{22}\\
  \end{pmatrix}
\end{equation}
and
\begin{equation}
 \hat{L}_z=
  \begin{pmatrix}
   0&0&0\\
   0&0&-2i\\
   0&2i&0\\
  \end{pmatrix}.
\end{equation}
Here, $\hat{\sigma}_0$ and $\hat{\sigma}_z$ are Pauli matrices and
$\lambda_{SOC}$ is the Ising-type SOC parameter. In monolayer NbSe$_2$,
$\lambda_{SOC}=0.0784$ eV.
$\hat{H}_{TNN}(\bm{k})$ includes the electron hoppings only among three
$d$-orbitals of Nb atoms, which are assumed up to third-nearest neighbor
sites as shown in Fig.~\ref{fig:5} (a).
Here, green, red and blue arrows indicate hopping vectors $\bm{R_i}$ $(i=1,2,\cdots,6)$ pointing to nearest-neighbor (n.n) sites, the vectors $\bm{\tilde{R_j}}$ $(j=1,2,\cdots,6)$ pointing to next n.n sites and the vectors $2\bm{R_i}$ pointing to third n.n sites, respectively.
We can find the matrix elements in the effective TBM Hamiltonian of
monolayer NbSe$_2$~\cite{Liu, Kim2021}: $V_0$, $V_1$, $V_2$, $V_{11}$, $V_{12}$ and $V_{22}$ as
\begin{equation}
  \begin{split}
    V_0=&\epsilon_1+2t_0(2\cos{\alpha}\cos{\beta}+\cos{2\alpha}) \\
    &+2r_0(2\cos{3\alpha}\cos{\beta}+\cos{2\beta})
    \\
    &+2u_0(2\cos{2\alpha}\cos{2\beta}+\cos{4\alpha}),
  \end{split}
\end{equation}
\begin{equation}
  \begin{split}
    \mathrm{Re}[V_1]=&-2\sqrt{3}t_2\sin{\alpha}\sin{\beta}+2(r_1+r_2)\sin{3\alpha}\sin{\beta}
    \\
    &-2\sqrt{3}u_2\sin{2\alpha}\sin{2\beta},
  \end{split}
\end{equation}
\begin{equation}
  \begin{split}
    \mathrm{Im}[V_1]=&2t_1\sin{\alpha}(2\cos{\alpha}+\cos{\beta})+2(r_1-r_2)\sin{3\alpha}\cos{\beta}
    \\
    &+2u_1\sin{2\alpha}(2\cos{2\alpha}+\cos{2\beta}),
  \end{split}
\end{equation}
\begin{equation}
  \begin{split}
    \mathrm{Re}[V_2]=&2t_2(\cos{2\alpha}-\cos{\alpha}\cos{\beta}) \\
    &-\frac{2}{\sqrt{3}}(r_1+r_2)(\cos{3\alpha}\cos{\beta}-\cos{2\beta})
    \\
    &+2u_2(\cos{4\alpha}-\cos{2\alpha}\cos{2\beta}),
  \end{split}
\end{equation}
\begin{equation}
  \begin{split}
    \mathrm{Im}[V_2]=&2\sqrt{3}t_1\cos{\alpha}\sin{\beta} \\
    &+\frac{2}{\sqrt{3}}(r_1-r_2)\sin{\beta}(\cos{3\alpha}+2\cos{\beta})
    \\
    &+2\sqrt{3}u_1\cos{2\alpha}\sin{2\beta},
  \end{split}
\end{equation}
\begin{equation}    
 \begin{split}
  V_{11}=&\epsilon_2+(t_{11}+3t_{22})\cos{\alpha}\cos{\beta}+2t_{11}\cos{2\alpha}
  \\
  &+4r_{11}\cos{3\alpha}\cos{\beta}+2(r_{11}+\sqrt{3}r_{12})\cos{2\beta}
  \\
  &+(u_{11}+3u_{22})\cos{2\alpha}\cos{2\beta}+2u_{11}\cos{4\alpha},
 \end{split}
\end{equation}
\begin{equation}
  \begin{split}
    \mathrm{Re}[V_{12}]=&\sqrt{3}(t_{22}-t_{11})\sin{\alpha}\sin{\beta}+4r_{12}\sin{3\alpha}\sin{\beta}
    \\
    &+\sqrt{3}(u_{22}-u_{11})\sin{2\alpha}\sin{2\beta},
  \end{split}
\end{equation}
\begin{equation}
  \begin{split}
    \mathrm{Im}[V_{12}]=&4t_{12}\sin{\alpha}(\cos{\alpha}-\cos{\beta})
    \\
    &+4u_{12}\sin{2\alpha}(\cos{2\alpha}-\cos{2\beta})
  \end{split}
\end{equation}
and
\begin{equation}
 \begin{split}
  V_{22}=&\epsilon_2+(3t_{11}+t_{22})\cos{\alpha}\cos{\beta}+2t_{22}\cos{2\alpha}
  \\
  &+2r_{11}(2\cos{3\alpha}\cos{\beta}+\cos{2\beta}) \\
  &+\frac{2}{\sqrt{3}}r_{12}(4\cos{3\alpha}\cos{\beta}-\cos{2\beta})
  \\
  &+(3u_{11}+u_{22})\cos{2\alpha}\cos{2\beta}+2u_{22}\cos{4\alpha}.
 \end{split}
\end{equation}
Here, $(\alpha, \beta)=(\frac{1}{2}k_xa, \frac{\sqrt{3}}{2}k_ya)$ and the lattice constant $a$ is $3.45$ \AA. The specific hopping parameters in this TBM can be given as
\begin{equation}
 E^{jj'}_{\mu \mu'}(\bm{R})=\braket{\psi^j_{\mu}(\bm{r})|\hat{H}(\bm{k})|\phi^{j'}_{\mu'}(\bm{r}-\bm{R})},
\end{equation}
where $|u_{n\bm{k}}\rangle \rightarrow \Ket{\phi^j_{\mu}}$ indicates an atomic orbital of Nb atom and in this paper, we consider $\Ket{\phi^1_1}=d_{z^2}$, $\Ket{\phi^2_1}=d_{xy}$ and $\Ket{\phi^2_2}=d_{x^2-y^2}$.
For example, we can express $t_0=E^{11}_{11}(\bm{R_1})$,
$t_1=E^{12}_{11}(\bm{R_1})$, $r_0=E^{11}_{11}(\bm{\tilde{R}_1})$,
$r_1=E^{12}_{11}(\bm{\tilde{R}_1})$, $u_0=E^{11}_{11}(2\bm{R_1})$ and
$u_1=E^{12}_{11}(2\bm{R_1})$.

Similarly, Hamiltonians of (AB-stacked) bilayer and (ABA-stacked) trilayer NbSe$_2$ can be obtained as
\begin{equation}
 \hat{H}_{bi}(\bm{k})=
  \begin{pmatrix}
   \hat{H}_{mono}(-\bm{k})&\hat{H}_{int}(\bm{k})\\
   \hat{H}^{\dag}_{int}(\bm{k})&\hat{H}_{mono}(\bm{k})\\
  \end{pmatrix}
\end{equation}
and
\begin{equation}
 \hat{H}_{tri}(\bm{k})=
  \begin{pmatrix}
   \hat{H}_{mono}(\bm{k})&\hat{H}_{int}(\bm{k})&0\\
   \hat{H}^{\dag}_{int}(\bm{k})&\hat{H}_{mono}(-\bm{k})&\hat{H}_{int}(\bm{k})\\
   0&\hat{H}^{\dag}_{int}(\bm{k})&\hat{H}_{mono}(\bm{k})\\
  \end{pmatrix},
\end{equation}
respectively.
In Fig.~\ref{fig:5} (b), interlayer hopping vector $\bm{R_{inter}}$ points $d$-orbital of upper layer to n.n sites of lower layer and the interlayer coupling Hamiltonian $\hat{H}_{int}(\bm{k})$ is considered as
\begin{equation}
 \hat{H}_{int}(\bm{k})=
  \begin{pmatrix}
   T_{01}&0&0\\
   0&T_{02}&0\\
   0&0&T_{02}\\
  \end{pmatrix},
\end{equation}
where $T_{01}$ and $T_{02}$ are fitted by using hopping parameters $t_{01}$ and $t_{02}$:
\begin{equation}
 T_{01}=3t_{01}+2t_{01}(2\cos{\alpha}\cos{\beta}+\cos{2\alpha}),
\end{equation}
\begin{equation}
 T_{02}=t_{02},
\end{equation}
respectively.
The details about fitted parameters for this TBM are
summarized in Table~\ref{tab:table2}.~\cite{He}

The energy band
structures and DOS are shown in Figs.~\ref{fig:5} (c), (d) and (e). Here, red,
blue and green lines indicate spin-up, spin-down and spin-degenerated
states, respectively.
The energy band structure of monolayer NbSe$_2$ is qualitatively similar
to heavily hole-doped monolayer MoS$_2$. Unlike monolayer MoS$_2$ which
shows semiconducting behavior, monolayer NbSe$_2$ is metallic, but a large
energy band gap between the partially filled valence bands and empty
conduction bands.
Also, the Ising-type SOC provides opposite spin
splitting at the valence band edges in K and K$^{\prime}$ points, and time-reversal
symmetry protection. In particular, the SOC makes the
spin splitting about $157$ meV at the K point.
Figure~\ref{fig:5} (d) shows that the spin splitting is absent in the
energy band structure of bilayer NbSe$_2$, but larger band splitting
appears at valence band in $\Gamma$ point because of the spacial inversion symmetry.
Figure~\ref{fig:5} (e) shows that trilayer NbSe$_2$ has spin splitting
bands owing to the broken crystal inversion symmetry as same as the
case of monolayer.
In gneral, the energy band structures of
NbSe$_2$ have spin-degeneracy along the $\Gamma$-M line,
which can be confirmed by looking at Fermi surface structure.
Also, this is because the Hamiltonians for up- and down-spin states have
the following properties:
$\hat{H}^{\uparrow}_{mono}(\bm{k})=(\hat{H}^{\downarrow}_{mono}(\bm{k}))^{\dag}$,
$\hat{H}^{\uparrow}_{bi}(\bm{k})=(\hat{H}^{\downarrow}_{bi}(\bm{k}))^{\dag}$
and
$\hat{H}^{\uparrow}_{tri}(\bm{k})=(\hat{H}^{\downarrow}_{tri}(\bm{k}))^{\dag}$,
respectively.
Since we have interest in the SHG, we discuss the nonlinear optical conductivity for SHG process ($\omega_1=\omega_2=\omega$).
Figure~\ref{fig:5} (c) includes the two SHG processes: (i) inter-band transition by two-photon absorption (green arrows) and (ii) inter-band transition even by one-photon absorption (red arrows).

Figure~\ref{fig:5} (f) shows Fermi surface of monolayer NbSe$_2$.
The surface has Fermi pockets centered at $\Gamma$, K and
K$^{\prime}$ points, which show the spin-splitting.
Also, because of the opposite spin-splitting around K and K$^{\prime}$
points, the energy band structure of monolayer NbSe$_2$ is anisotropic
with respect to $\Gamma$ point.
Figure~\ref{fig:5} (g) indicates Fermi surface of bilayer NbSe$_2$.
Unlike the case of monolayer, the surface has spin-degenerated Fermi
pockets owing to the crystal inversion symmetry.
Figure~\ref{fig:5} (h) shows Fermi surface of trilayer NbSe$_2$, which
has spin-splitting Fermi surface.
This Fermi surface can be understood by overlaying the spin-splitting Fermi surface
of monolayer NbSe$_2$ on the spin-degenerated Fermi surface of bilayer
NbSe$_2$.
We can mention that the spin dependence of the Fermi surface behaves differently for even-
and odd-number-layered NbSe$_2$.

\section{Berry curvature dipole in nonlinear optical conductivity}
We numerically calculate nonlinear optical spin and charge Hall
conductivities of SHG process for few-layered NbSe$_2$ based on an effective TBM.
In general, the 2nd order nonlinear optical conductivity $\sigma_{ijk}$ can be given as
\begin{widetext}
  \begin{equation}
    \begin{split}
      \sigma_{ijk} (\omega_1, \omega_2)=
  -\frac{\hbar^2e^2}{S}\sum_{\bm{k}}\sum_{nml}&\frac{1}{E_{ml}E_{ln}(E_{mn}-\hbar\omega_1-\hbar\omega_2-i\eta)}\Biggl[\frac{\braket{u_{n\bm{k}}|\hat{v}_j|u_{l\bm{k}}}\braket{u_{l\bm{k}}|\hat{v}_k|u_{m\bm{k}}}\braket{u_{m\bm{k}}|\hat{j}_i|u_{n\bm{k}}} f_{ml}}{E_{ml}-\hbar\omega_2-i\eta} \\
& -\frac{\braket{u_{n\bm{k}}|\hat{v}_k|u_{l\bm{k}}}\braket{u_{l\bm{k}}|\hat{v}_j|u_{m\bm{k}}}\braket{u_{m\bm{k}}|\hat{j}_i|u_{n\bm{k}}} f_{ln}}{E_{ln}-\hbar\omega_2-i\eta}\Biggl],
    \end{split}
    \label{eq:nonformwide}
  \end{equation}
\end{widetext}
where $i$($j$, $k$) indicates the direction $x$ or $y$.
In particular, the case where $ijk$ is $xyy$ and $yxx$ is called as Hall conductivity.
Also, $n$($m$, $l$) is the band index including spin degree
of freedom, $\Ket{u_{n\bm{k}}}$ is the eigen function with the
eigen energy $E_{n\bm{k}}$ and $f(E_{n\bm{k}})$ is Fermi-Dirac
distribution function.
$E_{ml}\equiv E_{m}-E_{l}$, $f_{ml}\equiv f(E_{m\bm{k}})-f(E_{l\bm{k}})$, $\eta$ is infinitesimally small real number and $S$ is the area of system.
$\hat{j}_i$ can be defined for spin and charge current operators, which can be written as $\hat{j}^{\rm{spin}}_i=\frac{1}{2}\{\frac{\hbar}{2}\hat{\sigma}_z\otimes \hat{I}_N, \hat{v}_i\}$ and $\hat{j}^{\rm{charge}}_i=\frac{1}{2}\{-e\hat{\sigma}_0\otimes \hat{I}_N, \hat{v}_i\}$, respectively.
Here, $\hat{I}_N$ is the $N\times N$ identity matrix and $N=3,6,9$ is
used for monolayer, bilayer and trilayer NbSe$_2$, respectively.
Moreover, $\hat{v}_{i}=\frac{1}{\hbar}\frac{\partial\hat{H}}{\partial i}$ is the group velocity operator.
We add the superscript ``$\rm{spin}$'' for the nonlinear optical spin conductivity $\sigma^{\rm{spin}}_{ijk} (\omega, \omega)$ in order to
distinguish its conductivity from the nonlinear optical charge conductivity $\sigma^{\rm{charge}}_{ijk} (\omega, \omega)$.
In case of $\omega_1=\omega_2=\omega$, the process is called as SHG.
In addition, in case of $\omega_1+\omega_2=\omega_3$, the process is called as SFG.

The 2nd order nonlinear optical conductivity expressed by Eq.~(\ref{eq:nonformwide})
can be separated into two inter-band processes using the following identity
\begin{widetext}
  \begin{equation}
    \frac{1}{(\hbar\omega_1+\hbar\omega_2+i\eta)(\hbar\omega_{mn}-\hbar\omega_1-\hbar\omega_2-i\eta)}=\frac{1}{(\hbar\omega_1+\hbar\omega_2+i\eta)\hbar\omega_{mn}}+\frac{1}{\hbar\omega_{mn}(\hbar\omega_{mn}-\hbar\omega_1-\hbar\omega_2-i\eta)},
  \end{equation}
\end{widetext}
where the first term leads to (i) optical transition between two bands $\sigma^{(2)}_{ijk}$ and the second term corresponds to (ii) optical transition involving three
bands $\sigma^{(3)}_{ijk}$.~\cite{Aversa1995, Passos2021} Namely, $\sigma_{ijk}$
can be decomposed into 
\begin{equation}
    \sigma_{ijk}(\omega_1, \omega_2)=\sigma^{(2)}_{ijk}(\omega_1, \omega_2)+\sigma^{(3)}_{ijk}(\omega_1, \omega_2).
\label{eq:2and3bandswide}
\end{equation}
In particular, $\sigma_{ijk}^{(2)}$ can be given as
\begin{widetext}
  \begin{equation}
    \begin{split}
      \sigma^{(2)}_{ijk}(\omega_1, \omega_2)=\frac{\hbar^2 e^2}{S}\sum_{\bm{k}}\sum_{nml}&\frac{1}{E_{ml}E_{ln}E_{mn}}\Biggl[\frac{\braket{u_{n\bm{k}}|\hat{v}_j|u_{l\bm{k}}}\braket{u_{l\bm{k}}|\hat{v}_k|u_{m\bm{k}}}\braket{u_{m\bm{k}}|\hat{j}_i|u_{n\bm{k}}} f_{ml}}{E_{ml}-\hbar\omega_2-i\eta} \\
      &-\frac{\braket{u_{n\bm{k}}|\hat{v}_k|u_{l\bm{k}}}\braket{u_{l\bm{k}}|\hat{v}_j|u_{m\bm{k}}}\braket{u_{m\bm{k}}|\hat{j}_i|u_{n\bm{k}}} f_{ln}}{E_{ln}-\hbar\omega_2-i\eta}\Biggl].
    \end{split}
    \label{eq:2bandslongwide}
  \end{equation}
\end{widetext}
Moreover, we can express $\sigma_{ijk}^{(2)}$ simply as inter-band transition between valence band $n$ and conduciton band $m$ using the following relations
\begin{widetext}
  \begin{equation}
    \hbar^2\sum_l\Biggl(\frac{\braket{u_{n\bm{k}}|\hat{v}_k|u_{l\bm{k}}}\braket{u_{l\bm{k}}|\hat{v}_j|u_{m\bm{k}}}}{E_{ln}E_{ml}}-\frac{\braket{u_{n\bm{k}}|\hat{v}_j|u_{l\bm{k}}}\braket{u_{l\bm{k}}|\hat{v}_k|u_{m\bm{k}}}}{E_{ln}E_{ml}}\Biggl)=\Biggl(\frac{\braket{u_{n\bm{k}}|\hat{v}_j|u_{m\bm{k}}}}{\omega_{nm}}\Biggl)_{;\bm{k^k}}-\Biggl(\frac{\braket{u_{n\bm{k}}|\hat{v}_k|u_{m\bm{k}}}}{\omega_{nm}}\Biggl)_{;\bm{k^j}},
  \end{equation}
where
  \begin{equation}
    \begin{split}
      \Biggl(\frac{\braket{u_{n\bm{k}}|\hat{v}_{\alpha}|u_{m\bm{k}}}}{\omega_{nm}}\Biggl)_{;\bm{k^{\beta}}}=&\frac{\braket{u_{n\bm{k}}|\hat{v}_{\beta}|u_{m\bm{k}}}\braket{u_{m\bm{k}}|\hat{\Delta}_{\alpha}|u_{n\bm{k}}}+\braket{u_{n\bm{k}}|\hat{v}_{\alpha}|u_{m\bm{k}}}\braket{u_{m\bm{k}}|\hat{\Delta}_{\beta}|u_{n\bm{k}}}}{i\omega_{nm}^2} \\
      &+\frac{i\hbar}{\omega_{nm}}\sum_l\Biggl(\frac{\braket{u_{n\bm{k}}|\hat{v}_{\beta}|u_{l\bm{k}}}\braket{u_{l\bm{k}}|\hat{v}_{\alpha}|u_{m\bm{k}}}}{E_{ln}}-\frac{\braket{u_{n\bm{k}}|\hat{v}_{\alpha}|u_{l\bm{k}}}\braket{u_{l\bm{k}}|\hat{v}_{\beta}|u_{m\bm{k}}}}{E_{ml}}\Biggl).
    \end{split}
  \end{equation}
\end{widetext}
Here, $\braket{u_{m\bm{k}}|\hat{\Delta}_j|u_{n\bm{k}}}\equiv \braket{u_{m\bm{k}}|\hat{v}_j|u_{m\bm{k}}}-\braket{u_{n\bm{k}}|\hat{v}_j|u_{n\bm{k}}}$ and $\alpha$($\beta$) is the direction $x$ or $y$.
In other word, $\sigma_{ijk}^{(2)}$ can be rewritten as
\begin{widetext}
  \begin{equation}
    \begin{split}
      \sigma^{(2)}_{ijk}(\omega_1, \omega_2)&=-\frac{i e^2}{S}\frac{1}{\hbar\omega_1+\hbar\omega_2+i\eta}\sum_{\bm{k}}\sum_{n}i\hbar^2\sum_{m\not=n}\frac{\braket{u_{n\bm{k}}|\hat{v}_j|u_{m\bm{k}}}\braket{u_{m\bm{k}}|\hat{v}_k|u_{n\bm{k}}}}{(E_{m\bm{k}}-E_{n\bm{k}})^2}\braket{u_{n\bm{k}}|\hat{j}_i|u_{n\bm{k}}} \frac{f_{mn}}{E_{mn}-\hbar\omega_2-i\eta} \\
    &=-\frac{i e^2}{S}\frac{1}{\hbar\omega_1+\hbar\omega_2+i\eta}\sum_{\bm{k}}\sum_{n}\Omega_{n}(\bm{k})\braket{u_{n\bm{k}}|\hat{j}_i|u_{n\bm{k}}} \frac{f_{mn}}{E_{mn}-\hbar\omega_2-i\eta},
    \end{split}
    \label{eq:2bandsnotfinwide}
  \end{equation}
\end{widetext}
where $\Omega_{n}(\bm{k})$ is the Berry curvature, i.e.,
\begin{equation}
    \Omega_{n}(\bm{k})=i\hbar^2\sum_{m\not=n}\frac{\braket{u_{n\bm{k}}|\hat{v}_j|u_{m\bm{k}}}\braket{u_{m\bm{k}}|\hat{v}_k|u_{n\bm{k}}}}{(E_{m\bm{k}}-E_{n\bm{k}})^2}.
\label{eq:berrycurvwide}
\end{equation}
When we consider the inter-band transition between two bands around Fermi surface, $\hbar\omega_2$ in Eq.~(\ref{eq:2bandsnotfinwide}) is infinitesimally zero, i.e.,
\begin{equation}
  \begin{split}
    \sigma^{(2)}_{ijk}(\omega_1, \omega_2\rightarrow 0)&=-\frac{i e^2}{S}\frac{1}{\hbar\omega_1+i\eta}\sum_{\bm{k}}\sum_{n}\Omega_{n}(\bm{k})\braket{u_{n\bm{k}}|\hat{j}_i|u_{n\bm{k}}} \frac{f_{mn}}{E_{mn}} \\
    &=-\frac{i e^2}{S}\frac{1}{\hbar\omega_1+i\eta}\sum_{\bm{k}}\sum_{n}\Omega_n(\bm{k})\braket{u_{n\bm{k}}|\hat{j}_i|u_{n\bm{k}}} \frac{\partial{f_{n\bm{k}}}}{\partial{E_{n\bm{k}}}} \\
    &=-\frac{i e^2}{S}\frac{1}{\hbar\omega_1+i\eta}D_i,
  \end{split}
\label{eq:2bandswide}
\end{equation}
where $D_i$ is the Berry curvature dipole~\cite{Zeng2021} and we have used an approximation to Fermi-Dirac distribution
\begin{equation}
  f_{mn} = f_{m\bm{k}}-f_{n\bm{k}}\approx \frac{\partial{f_{n\bm{k}}}}{\partial{E_{n\bm{k}}}}(E_{m\bm{k}}-E_{n\bm{k}}).
\end{equation}
It should be noted that $\sigma^{(2)}_{ijk}$ includes $\Omega_n(\bm{k})$, velocity and energy derivative of Fermi-Dirac distribution function.
Also, $\sigma_{ijk}^{(3)}$ can be given as
\begin{widetext}
  \begin{equation}
    \begin{split}
      \sigma^{(3)}_{ijk}(\omega_1, \omega_2)=\frac{\hbar^2 e^2}{S}\sum_{\bm{k}}\sum_{nml}&\frac{\hbar\omega_1+\hbar\omega_2+i\eta}{E_{ml}E_{ln}E_{mn}(E_{mn}-\hbar\omega_1-\hbar\omega_2-i\eta)}\Biggl[\frac{\braket{u_{n\bm{k}}|\hat{v}_j|u_{l\bm{k}}}\braket{u_{l\bm{k}}|\hat{v}_k|u_{m\bm{k}}}\braket{u_{m\bm{k}}|\hat{j}_i|u_{n\bm{k}}} f_{ml}}{E_{ml}-\hbar\omega_2-i\eta} \\
      &-\frac{\braket{u_{n\bm{k}}|\hat{v}_k|u_{l\bm{k}}}\braket{u_{l\bm{k}}|\hat{v}_j|u_{m\bm{k}}}\braket{u_{m\bm{k}}|\hat{j}_i|u_{n\bm{k}}} f_{ln}}{E_{ln}-\hbar\omega_2-i\eta}\Biggl],
    \end{split}
    \label{eq:3bandswide}
  \end{equation}
\end{widetext}
where $\sigma_{ijk}^{(3)}$ includes three bands, i.e., a valence band $n$, an intermediate band $l$ and a conduction band $m$.

\section{Integrands of nonlinear optical spin and charge Hall conductivities}
\begin{figure*}[t]
  \begin{center}
    \includegraphics[width=1.0\textwidth]{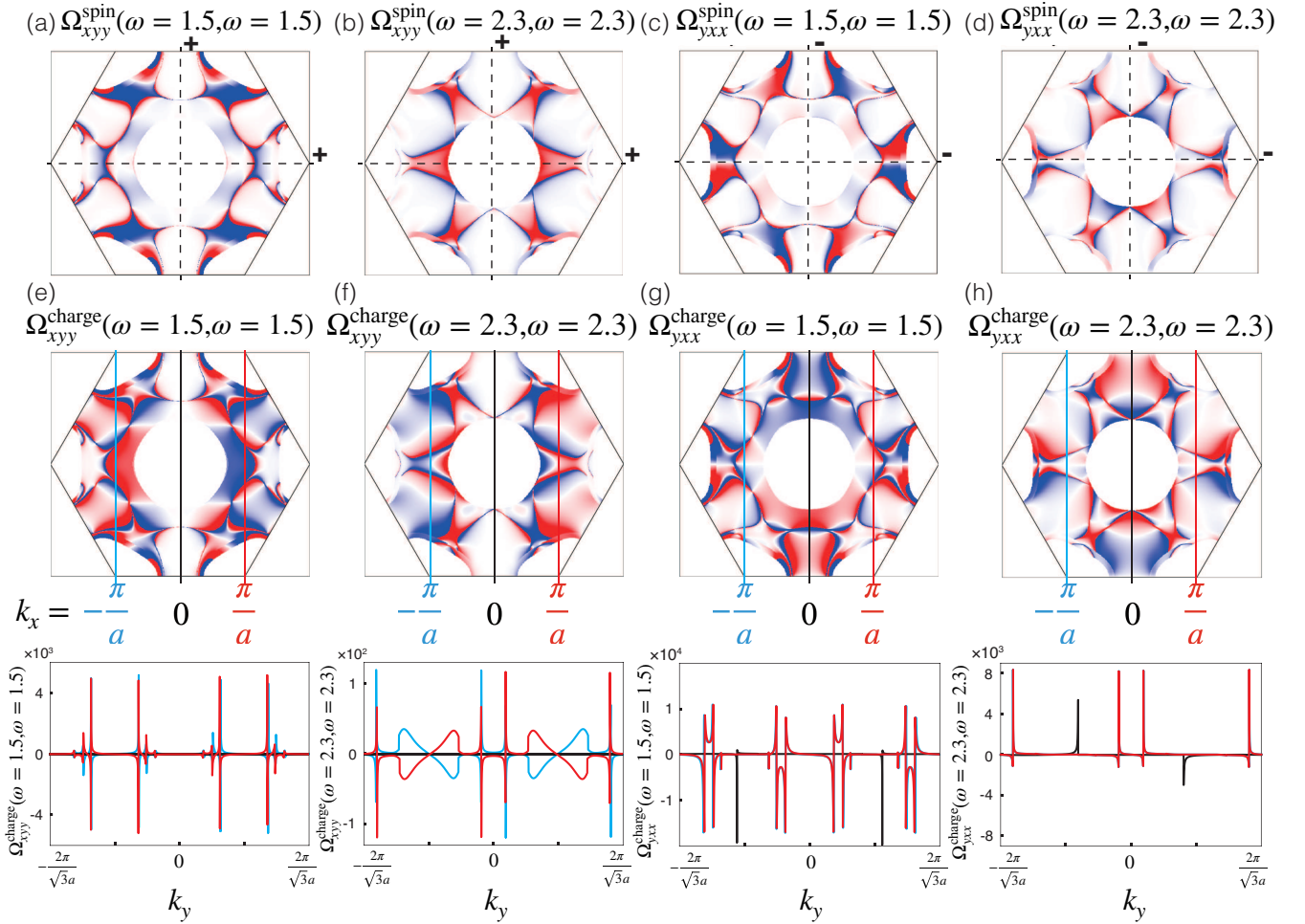}
    \caption{Contour plots of integrands for nonlinear optical spin Hall
   conductivities of monolayer NbSe$_2$ in the 1st BZ: 
      (a) $\Omega^{\rm{spin}}_{xyy}(\omega=1.5, \omega=1.5)$, (b)
      $\Omega^{\rm{spin}}_{xyy}(\omega=2.3, \omega=2.3)$, (c)
      $\Omega^{\rm{spin}}_{yxx}(\omega=1.5, \omega=1.5)$ and (d) $\Omega^{\rm{spin}}_{yxx}(\omega=2.3, \omega=2.3)$.
      Same plots of integrands for nonlinear optical charge Hall
   conductivities of monolayer NbSe$_2$: (e) $\Omega^{\rm{charge}}_{xyy}(\omega=1.5, \omega=1.5)$, (f)
      $\Omega^{\rm{charge}}_{xyy}(\omega=2.3, \omega=2.3)$, (g) $\Omega^{\rm{charge}}_{yxx}(\omega=1.5, \omega=1.5)$ and (h) $\Omega^{\rm{charge}}_{yxx}(\omega=2.3, \omega=2.3)$.}
    \label{fig:6}
  \end{center}
\end{figure*}

We calculate integrands of nonlinear optical spin Hall conductivities
$\Omega^{\rm{spin}}_{xyy} (\omega, \omega, \bm{k})$ and
$\Omega^{\rm{spin}}_{yxx} (\omega, \omega, \bm{k})$ of few-layered NbSe$_2$.
According to Fig.~\ref{fig:2}, the nonlinear optical spin Hall conductivity of even-number-layered
NbSe$_2$ is absent owing to Neumann's principle, but that of
odd-number-layered NbSe$_2$ has finite values.
Here, we especially discuss the case of monolayer.
Figures~\ref{fig:6} (a) and (b) show the contour plots of
$\Omega^{\rm{spin}}_{xyy} (\omega, \omega, \bm{k})$ at $\hbar\omega=1.5$ and
$2.3$ eV, respectively. 
Under light irradiation, $\Omega^{\rm{spin}}_{xyy} (\omega, \omega, \bm{k})$ has
even-parity with respect to $k_x$ and $k_y$ axes.
Thus, the $k$-integration of $\Omega^{\rm{spin}}_{xyy} (\omega, \omega, \bm{k})$ over
1st BZ becomes nonvanishing and reproduces the result of
Fig.~\ref{fig:2} (a).
On the other hand, Figures~\ref{fig:6} (c) and (d) show 
$\Omega^{\rm{spin}}_{yxx} (\omega, \omega, \bm{k})$ under light irradiation of
$\hbar\omega=1.5$ and $2.3$ eV, which has
odd-parity for $k_x$ and $k_y$ axes.
Owing to the antisymmetric properties of 
$\Omega^{\rm{spin}}_{yxx} (\omega, \omega, \bm{k})$ for monolayer NbSe$_2$, 
their $k$-integration over 1st BZ has
finite value and is consistent with the result of
Sec.~\ref{sec:level3}.
Thus, the nonlinear optical spin Hall current can be generated in
$x$-direction for monolayer NbSe$_2$ by irradiating $y$-polarized
light.

Next we discuss $\Omega^{\rm{charge}}_{xyy} (\omega, \omega, \bm{k})$ and
$\Omega^{\rm{charge}}_{yxx} (\omega, \omega, \bm{k})$ of odd-number-layered
NbSe$_2$.
Owing to the crystal symmetry,  Neumann's principle
illustrates that the nonlinear optical charge Hall current is absent (generated) 
in even (odd)-number-layered NbSe$_2$. 
Figures~\ref{fig:6} (e) and (f) show the contour plots of 
$\Omega^{\rm{charge}}_{xyy} (\omega, \omega, \bm{k})$ for monolayer NbSe$_2$
 under light irradiation of $\hbar\omega=1.5$ and $2.3$ eV,
 respectively. 
In addition, below the each contour plot, the sliced
$\Omega^{\rm{charge}}_{xyy} (\omega, \omega, \bm{k})$ at
$k_x=-\frac{\pi}{a}$, $0$ and $\frac{\pi}{a}$ is shown. 
Here, blue, black and red lines indicate $\Omega^{\rm{charge}}_{xyy} (\omega, \omega \bm{k})$ at $k_x=-\frac{\pi}{a}$, $0$ and $\frac{\pi}{a}$,
respectively. 
$\Omega^{\rm{charge}}_{xyy} (\omega, \omega, \bm{k})$ of $k_x=-\frac{\pi}{a}$
has opposite sign to that of $k_x=\frac{\pi}{a}$. Therefore, 
the $k$-integration of $\Omega^{\rm{charge}}_{xyy} (\omega, \omega, \bm{k})$
over 1st BZ becomes identically zero, which is consistent with
the result of Sec.~\ref{sec:level3}.
Figures~\ref{fig:6} (g) and (h) show
the contour plots of $\Omega^{\rm{charge}}_{yxx} (\omega, \omega, \bm{k})$ 
in the 1st BZ under the light irradiation of $\hbar\omega=1.5$ and $2.3$ eV, respectively.
Also, we show the sliced $\Omega^{\rm{charge}}_{yxx} (\omega, \omega, \bm{k})$ 
at $k_x=-\frac{\pi}{a}$, $0$ and $\frac{\pi}{a}$.
Because $\Omega^{\rm{charge}}_{yxx} (\omega, \omega, \bm{k})$ of $k_x=-\frac{\pi}{a}$
has same sign to that of $k_x=\frac{\pi}{a}$, it is clear that the
$k$-integration of $\Omega^{\rm{charge}}_{yxx} (\omega, \omega, \bm{k})$ over
1st BZ becomes finite, which reproduces the result of Fig.~\ref{fig:2}
(d).
Thus, the nonlinear optical charge Hall current can be generated in
$y$-direction for monolayer NbSe$_2$ by irradiating $x$-polarized
light.

\section{Energy band structures of bilayer NbSe$_2$ with applied
  electric fields}
\begin{figure*}[t]
  \begin{center}
    \includegraphics[width=0.95\textwidth]{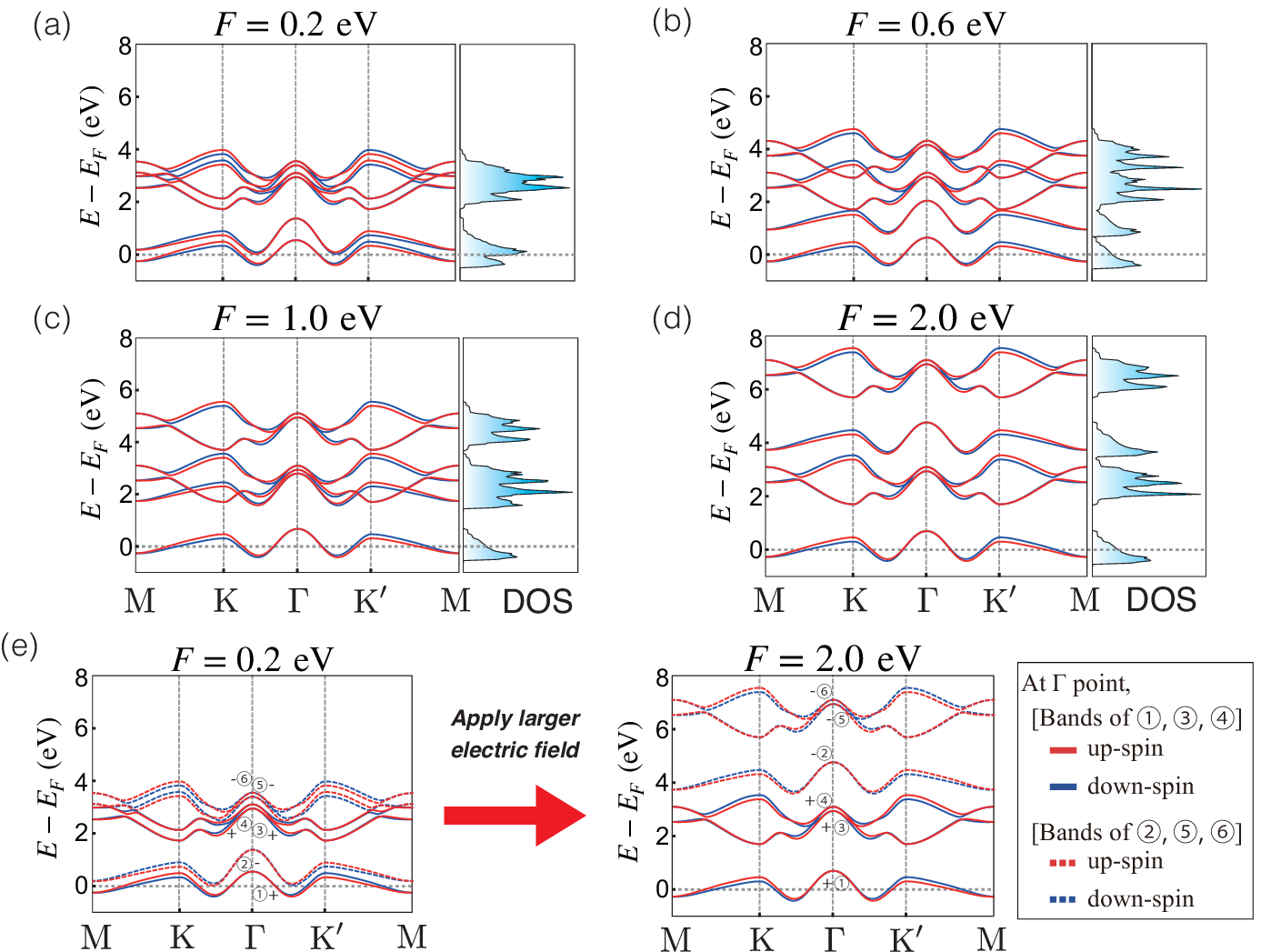}
   \caption{Energy band structures of bilayer NbSe$_2$ with applied
     electric fields of (a) $F=0.2$, (b) $0.6$,
     (c) $1.0$, (d) $2.0$ eV, respectively.
     Fermi levels are set to (a) $-0.1370$, (b) $-0.5165$, (c)
     $-0.9113$ and (d) $-1.9074$ eV, respectively.
     (e) Bilayer NbSe$_2$ with applied electric field includes 6
     energy bands $\textcircled{\scriptsize1}$ to
     $\textcircled{\scriptsize6}$.
     At $\Gamma$ point, the energy bands of
     $\textcircled{\scriptsize1}$, $\textcircled{\scriptsize3}$ and
     $\textcircled{\scriptsize4}$ are shown by solid lines.
     The energy bands of $\textcircled{\scriptsize2}$, $\textcircled{\scriptsize5}$ and $\textcircled{\scriptsize6}$ are shown by dashed lines.}
    \label{fig:7}
  \end{center}
\end{figure*}

Figures~\ref{fig:7} (a), (b), (c) and (d) show the energy band
structures and DOS of bilayer NbSe$_2$ for several different electric
fields. 
Here, red and blue lines show spin-up and spin-down states,
respectively.
Owing to the broken inversion symmetry in bilayer NbSe$_2$ with
applied electric fields, the energy band structures have larger spin
splitting at the valence band edges in K and K$^{\prime}$ points as
same as the case of odd-number-layered NbSe$_2$.
In this paper, we provide the energy band structures with applied
electric fields of $F=0.2$, $0.6$, $1.0$ and $2.0$ eV, respectively.
Fermi levels are set to $-0.1370$, $-0.5165$, $-0.9113$ and
$-1.9074$ eV for the applied electric fields of $F=0.2$, $0.6$, $1.0$ and $2.0$ eV,
respectively.

Figure~\ref{fig:7} (e) indicates the energy bands can move by applying
larger electric field in bilayer NbSe$_2$.
Here, the energy bands of solid lines ($\textcircled{\scriptsize1}$, $\textcircled{\scriptsize3}$ and $\textcircled{\scriptsize4}$)
have positive parity between upper and
lower layers of bilayer NbSe$_2$, i.e., bonding molecular orbitals
between the two layers.
On the other hand, the energy bands of dashed lines ($\textcircled{\scriptsize2}$,
$\textcircled{\scriptsize5}$ and $\textcircled{\scriptsize6}$)
have negative parity between layers, i.e. anti-bonding configuration. 
With increase of electric field,
the energy bands with negative (positive) parity shift up (down) to
higher (lower) energy. 

\section{Bilayer NbSe$_2$ with each layer having a different Fermi energy}
\begin{figure*}[t]
  \begin{center}
    \includegraphics[width=0.95\textwidth]{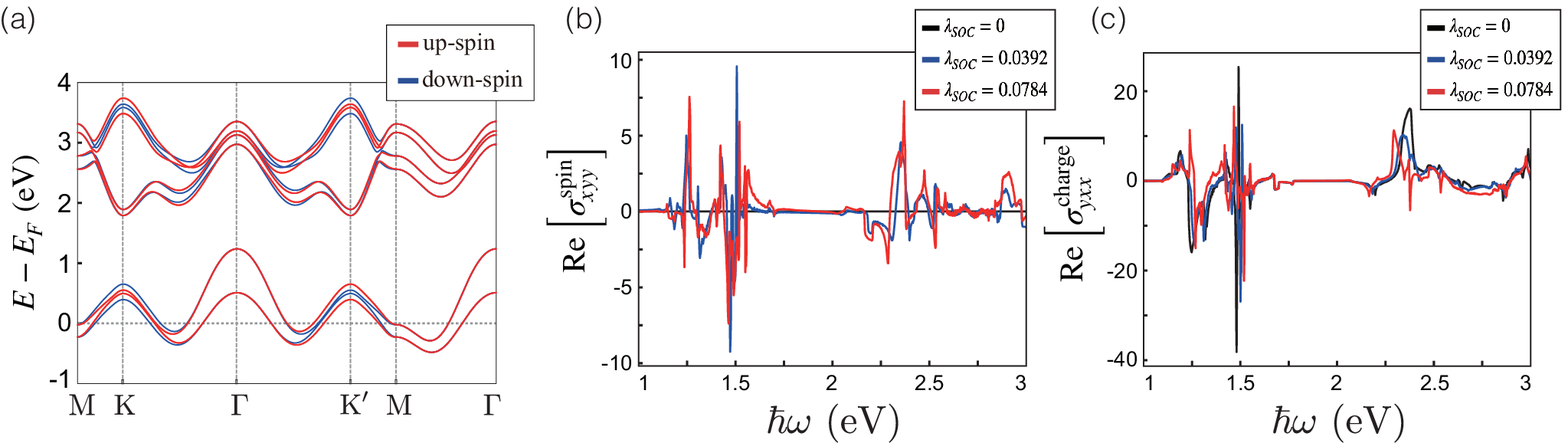}
    \caption{(a) Energy band structure of bilayer NbSe$_2$ with each layer having a different Fermi energy. (b) Real part of nonlinear optical spin Hall conductivity $\mathrm{Re}\left[\sigma^{\rm{spin}}_{xyy}(\omega, \omega)\right]$ of decoupled bilayer NbSe$_2$. (c) Real part of nonlinear optical charge Hall conductivity $\mathrm{Re}\left[\sigma^{\rm{charge}}_{yxx}(\omega, \omega)\right]$ of decoupled bilayer NbSe$_2$. The units of $\mathrm{Re}\left[\sigma^{\rm{spin}}_{xyy}(\omega, \omega)\right]$ and $\mathrm{Re}\left[\sigma^{\rm{charge}}_{yxx}(\omega, \omega)\right]$ are $e^2$ and $e^3/\hbar$, respectively.}
    \label{fig:8}
  \end{center}
\end{figure*}

In this section, we consider the nonlinear optical spin and charge
conductivities of bilayer NbSe$_2$ with each layer having a different
Fermi energy, i.e. decoupled bilayer NbSe$_2$. 
Figure~\ref{fig:8} (a) shows the energy band structure of decoupled bilayer
NbSe$_2$, where Fermi energy of the upper layer is $E_F=0$ 
and that of the lower layer is $0.1$ eV, respectively.
Here, red and blue lines show spin-up and spin-down states, respectively.
Since the upper and lower layers have the different Fermi energy, 
the energy band structure has spin splitting at K and K$^{\prime}$ points as same as the case of odd-number-layered NbSe$_2$.

Figures~\ref{fig:8} (b) and (c) show the real parts of nonlinear
optical spin and charge Hall conductivities
$\mathrm{Re}\left[\sigma^{\rm{spin}}_{xyy} (\omega, \omega)\right]$ and $\mathrm{Re}\left[\sigma^{\rm{charge}}_{yxx} (\omega, \omega)\right]$ of
decoupled bilayer NbSe$_2$, respectively.
Here, both $\mathrm{Re}\left[\sigma^{\rm{spin}}_{xyy} (\omega, \omega)\right]$ 
and $\mathrm{Re}\left[\sigma^{\rm{charge}}_{yxx}(\omega, \omega)\right]$ 
are considerd as the case of SHG process and have Ising-type SOC parameter $\lambda_{SOC}=0.0784$ eV.
Also, the cases for $\lambda_{SOC}=0.0392$ and $0$ eV are plotted for the comparison. 
Because of the Fermi energy imbalance between upper and lower layers,
the nonlinear optical spin and charge Hall conductivities become finite 
even in the bilayer NbSe$_2$, where
several peaks appear below $1.5$ and above $2.2$ eV. 
The details of the imaginary parts of nonlinear optical spin and charge
Hall conductivities are shown in Supplementary Material.

\nocite{*}
\bibliography{reference}
\end{document}